\documentclass[12pt,a4paper]{article}

\usepackage{graphicx}

\newcommand{\ordo}{\mathcal{O}}
\newcommand{\tree}
{\hspace*{1ex}{\raisebox{1.7ex}{\scriptsize $\circ$}\hspace{-2ex}}}
\newcommand{\treel}
{\hspace*{1.0ex}{\raisebox{1.7ex}{\scriptsize $\circ$}\hspace{-1.2ex}}}

\newcommand{\cF}{\mathcal{F}}
\newcommand{\cN}{\mathcal{N}}
\newcommand{\mpi}{M_{\pi}}
\newcommand{\fpi}{F_{\pi}}
\newcommand{\Mpi}{M_{\pi}}
\newcommand{\Fpi}{F_{\pi}}
\newcommand{\mk}{M_K}
\newcommand{\fk}{F_K}
\newcommand{\me}{M_\eta}
\newcommand{\fe}{F_\eta}

\newcommand{\ovr}{\over}
\newcommand{\til}{\tilde}

\newcommand{\<}{\langle}
\renewcommand{\>}{\rangle}

\newcommand{\et}{\eta}

\newcommand{\la}{\lambda}

\newcommand{\si}{\sigma}

\newcommand{\ph}{\phi}

\newcommand{\om}{\omega}

\newcommand{\be}{\begin{equation}}
\newcommand{\ee}{\end{equation}}
\newcommand{\beq}{\begin{equation}}
\newcommand{\eeq}{\end{equation}}
\newcommand{\bea}{\begin{eqnarray}}
\newcommand{\eea}{\end{eqnarray}}
\newcommand{\bdm}{\begin{displaymath}}
\newcommand{\edm}{\end{displaymath}}

\newcommand{\mr}{\mathrm}
\newcommand{\mb}{\mathbf}
\newcommand{\Nf}{N_{\!f\,}}
\newcommand{\MeV}{\,\mr{MeV}}
\newcommand{\GeV}{\,\mr{GeV}}

\newcommand{\fm}{\,\mr{fm}}

\newcommand{\ri}{\mr{i}}

\newcommand{\lb}{\bar\ell}
\newcommand{\rt}{\tilde r}

\newcommand{\nn}{\nonumber\\}
\def\fs{\;\;.}

\begin{document}

\title{\LARGE\bf Finite volume effects\\
for meson masses and decay constants\\[8mm]}

\author{{\bf Gilberto Colangelo}, {\bf Stephan D\"urr} and
{\bf Christoph Haefeli}\\[2mm]
{\small Institut f\"ur Theoretische Physik, Universit\"at Bern,
Sidlerstr. 5, 3012 Bern, Switzerland}}

\date{14 March 2005} 
\maketitle

\begin{abstract}
We present a detailed numerical study of finite volume effects for masses and
decay constants of the octet of pseudoscalar mesons. For this analysis we use
chiral perturbation theory and asymptotic formulae {\em \`a la} L\"uscher and
propose an extension of the latter beyond the leading exponential term. We
argue that such a formula, which is exact at the one-loop level, gives the
numerically dominant part at two loops and beyond. Finally, we discuss the
possibility to determine low energy constants from 
the finite volume dependence of masses and decay constants.
\end{abstract}

\thispagestyle{empty}
\setcounter{page}{0}
\clearpage

\tableofcontents
\thispagestyle{empty}
\setcounter{page}{0}
\clearpage


\section{Introduction}


In lattice QCD the determination of the mass and decay constant of the
lowest-lying state with a given set of quantum numbers 
is entering the high-precision era.
Even in the fully unquenched case (i.e.\ with sea- and valence-quarks being
degenerate) the pion mass can be measured, for fixed bare parameters, with
an accuracy at the percent level, and future progress towards the permille
level is anticipated.  However, to make contact with the real world, three
extrapolations are needed.  These are $(i)$ the continuum extrapolation,
$(ii)$ the infinite volume extrapolation, and $(iii)$ the chiral
extrapolation.
In each case there is considerable help and an analytical 
guideline from an effective field theory framework.

This article is concerned with the extrapolation to infinite volume, where
the situation is particularly favorable (for a compact review of the recent
literature see \cite{Colangelo:2004sc}).  As shown by Gasser and Leutwyler,
chiral symmetry imposes strong constraints on the dynamics at low energy in
QCD, even if the system is enclosed in a finite box.  Accordingly, Chiral
Perturbation Theory (ChPT) may be adapted to the finite volume case
\cite{Gasser:1986vb,Gasser:1987ah,Gasser:1987zq}.
In this framework finite volume effects can be taken
systematically into account, in a perturbative loop expansion.  If the
spatial volume $L^3$ is large enough internal degrees of freedom of the
particle of interest play no role as far as finite volume effects are
concerned and these are exclusively due to pion loops. This means that
they first appear at next-to-leading order (NLO), i.e.\ at $O(p^4)$ in the
chiral counting.  Another consequence is that they are exponentially small
in the \emph{pion} mass for any particle that couples to the pion field:
the effect behaves like $e^{-\Mpi L}$ for pions, kaons, etas and nuclei.
To date, a number of finite volume calculations have been performed at
one-loop order. In the nucleon sector the corrections to the mass
\cite{AliKhan:2003cu,Beane:2004tw}, the magnetic moment \cite{Beane:2004tw}
and the baryon axial charge \cite{Beane:2004rf} have been worked out.  In
the meson sector the original calculation for $\Mpi(L)$ and $\Fpi(L)$
\cite{Gasser:1986vb} has been extended to the quenched case
\cite{Sharpe:1992ft}. The same quantities have also been analyzed in a
quark-meson model \cite{Braun:2004yk}.
More recently, the finite volume shifts for $F_K(L)$
and $B_K(L)$ have been given in Ref.\,\cite{Becirevic:2003wk} and the
extension to heavy-meson chiral perturbation theory has been described
\cite{Arndt:2004bg}. No full two-loop calculation of finite volume effects
has appeared yet.

For some observables the L\"uscher formula represents a convenient and
powerful alternative \cite{Luscher:1985dn}.  It allows to estimate
subleading (in the chiral counting) finite volume effects for the mass of a
particle with less effort, while sticking to the leading order in an
expansion in powers of $e^{-\Mpi L}$. The formula gives the finite volume
shift $M_P(L)\!-\!M_P$ of a particle $P$ in terms of the \emph{infinite}
volume $\pi P$ forward scattering amplitude in the unphysical (Euclidean)
region. For this amplitude the ChPT expression at a certain loop order is
used. The approach via the L\"uscher formula is economical, since only a
chiral calculation in infinite volume is needed, and the loop in finite
volume comes for free.  Applications of the L\"uscher formula to the mass
of the nucleon \cite{AliKhan:2003cu,Koma:2004wz} and the pion
\cite{Colangelo:2003hf} have been worked out.  In the latter publication
several orders in the ChPT input for the $\pi\pi$ scattering amplitude were
compared and it was found that in a certain range of $\Mpi$ and $L$
subleading effects (in the chiral counting) can be large with respect to
the leading contributions.  An extension of the L\"uscher formula has been
constructed for pseudoscalar decay constants \cite{Colangelo:2004xr}.

This article presents a resummed version of the L\"uscher formulae for
masses and decay constants (this has been briefly discussed by one of us in
\cite{Colangelo:2004sc}), where the terms neglected in the large-$L$
expansion are $\sim\!e^{-(\sqrt{3}+1)/\sqrt{2}\cdot\Mpi L}$, rather than
$e^{-\sqrt{2}\Mpi L}$.  We proceed with the asymptotic expression for
$\Mpi(L)\!-\Mpi$ to 3-loop order, and for $\Fpi(L)\!-\!\Fpi,
M_K(L)\!-\!M_K, F_K(L)\!-\!F_K$ to 2-loop order, by using the available
knowledge in the literature on the scattering amplitude or axial-vector
matrix element.
In all cases, the result may be given in
a compact formula that does not involve any numerical integration.


\section{Finite volume effects}


Lattice calculations are necessarily done in a finite 4D volume which
acts as an IR-cutoff. Typically a $L^3\!\times\!T$ geometry with periodic
boundary conditions in all directions is chosen with $T\!\gg\!L$ and both
large compared to the inverse temperature of the QCD phase transition or
crossover. In order to determine the mass of a particle one considers a
correlator of two properly chosen interpolating fields
\beq
 C(t)=\int d^3 x\;\<\ph(x)\ph(0)\>\;e^{\ri \mb{px}}
\; \; \stackrel{T \to \infty}{\longrightarrow} \; \; \sum_{n=0}^\infty
c_n\,e^{-E_nt} 
\eeq
and tries to determine the energy levels (typically the lowest) at
zero spatial momentum, $M_n(L)=E_n(\mb{p}=0)$. 
In the following, we will be concerned with the shift $M(L)-M$, where
$M\!\equiv\!M(L\!=\!\infty)$, for the groundstate ($n\!=\!0$) due to the
finite 3D volume $L^3$.
Decay constants or, more generally, matrix elements suffer from analogous
shifts at finite spatial volume.
For not-too-small box length $L$ these shifts can be calculated analytically,
thus offering a means to correct lattice data for this systematic effect.
We now give a brief outline of the two main frameworks for such a calculation.
A comment on cut-off effects and their interplay with finite volume effects
is given in appendix~\ref{app:cutoff}.

\subsection{ChPT in finite volume}

In QCD with light flavors the physics in the infrared region is controlled
by chiral symmetry. 
As shown by Gasser and Leutwyler this still holds true if the system is
enclosed in a finite box $L^3\!\times\!T$, provided both $L$ and $T$ are
large enough that chiral symmetry is not restored \cite{Gasser:1987zq}.
They have shown that in an isotropic box with periodic boundary conditions
for the meson fields the finite-volume dependence comes in exclusively
through the propagators.
The latter becomes periodic in all spatial directions, and in the limit 
$T \to \infty$ can be written as follows
\beq
G(x^0,\mb{x})=\sum_{\mb{n}}G_0(x^0,\mb{x}+\mb{n}L)
\eeq
which is equivalent to replacing the integration over the spatial part of
the momenta by a sum over multiples of $2\pi/L$.
In other words, with periodic boundary conditions the Lagrangian remains the
\emph{same} as in infinite volume.

The expansion parameters in ChPT are 
\beq
 {p\ovr4\pi\Fpi} \qquad  , \qquad{\Mpi\ovr4\pi\Fpi}
\eeq
and the theory can be meaningfully applied only if both are small. In a
finite volume spatial momenta are discretized: $\mb{p}=2 \pi \mb{n} /L$
with $\mb{n}$ a vector of integers. Therefore one can have ``small''
nonzero momenta and apply ChPT only if the condition
\beq
L\!\gg\!{1\ovr2\Fpi}\!\sim\!1\fm
\label{xpt_cond1}
\eeq
is satisfied. 
A priori there is no way to say how much $L$ has to be in excess of 1 fm.
As a guideline we observe that the lowest non-trivial momentum in a 1 fm
box is $1.2\GeV$, which is certainly beyond the realm of ChPT.
Note, finally, that unlike $\Fpi L$ the combination $\Mpi L$ is not
constrained.
Both $\Mpi L\!\ll\!1$ and $\Mpi L\!\gg\!1$ are acceptable
\cite{Gasser:1986vb,Gasser:1987ah,Gasser:1987zq,betterways},
but they imply different ways to organize the chiral series,
\bea
\mpi L \gg 1 \quad \leftrightarrow \quad \mbox{``$p$-expansion''}\,\,
\label{xpt_mom}
\\
\mpi L \ll 1 \quad \leftrightarrow \quad \mbox{``$\epsilon$-expansion''}
\;.
\label{xpt_eps}
\eea
Here we shall restrict ourselves to the former case, where the chiral
counting is
\beq
\Mpi^2\sim m\sim O(p^2), \qquad
1/L\sim p\sim O(p)\;.
\eeq
With this setup Gasser and Leutwyler calculated the mass and decay constant
shift in a theory with $\Nf^2\!-\!1$ degenerate pseudo-Goldstone bosons, and
obtained \cite{Gasser:1986vb}
\bea
\Mpi(L)&=&\Mpi
\left[1+{1\ovr2\Nf}\xi_\pi\,\tilde g_1(\lambda_\pi)+O(\xi_\pi^2)\right]
\label{mpi}
\\
\Fpi(L)&=&\;\Fpi
\left[1-\;{\Nf\over2}\,\xi_\pi\,\tilde g_1(\lambda_\pi)+O(\xi_\pi^2)\right]
\;.
\label{fpi}
\eea
Here we have introduced the abbreviations (note the $\Fpi$ in the denominator
for all $P$)
\bea
\xi_P&\equiv&{M_P^2\ovr(4\pi\Fpi)^2}
\label{def_xi_p}
\\
\lambda_P&\equiv&M_P L
\label{def_la_p}
\eea
for $P\!=\!\pi$ (and $P\!=\!K,\eta$ will be used below)
as well as the modified
\footnote{Our $\til g_1$ relates to $g_1$ of \cite{Gasser:1986vb} via
$\til g_1(\la_\pi)=(4\pi/\Mpi)^2\!\cdot\!g_1(\Mpi,\beta\!=\!\infty,L)$ and is
a dimensionless function.}
shape function
\beq
\tilde g_1(x) = \sum_{n=1}^\infty
{4 m(n)\ovr\sqrt{n}\,x}\; K_1(\sqrt{n}\,x)
\label{g1til}
\eeq
where $K_1$ is a Bessel function of the second kind and the multiplicities
$m(n)$ have been given in \cite{Colangelo:2003hf}, but for convenience we
reproduce them in Tab.\,\ref{tab:multiplicities}.
Given the asymptotic expansion $K_1(z)\!\sim\!\sqrt{\pi/(2z)}e^{-z}$, it is
clear that in the $p$-regime eqns.\,(\ref{mpi}, \ref{fpi}, \ref{g1til})
represent quickly converging expressions.
Several observables have been worked out at one-loop order
\cite{AliKhan:2003cu,Beane:2004tw,Beane:2004rf,Sharpe:1992ft,Becirevic:2003wk,
Arndt:2004bg,Descotes-Genon:2004iu,Detmold:2005pt}, but to date no two-loop
result obtained in this setup has appeared.

\begin{table}[t]
\begin{tabular}{|l|rrrrrrrrrrrrrrrrrrrr|}
\hline
$n$&1&2&3&4&5&6&7&8&9&10&11&12&13&14&15&16&17&18&19&20\\
$m(n)$&6&12&8&6&24&24&0&12&30&24&24&8&24&48&0&6&48&36&24&24\\
\hline
\end{tabular}
\caption{The multiplicities $m(n)$ in (\ref{g1til}) for $n\leq20$.}
\label{tab:multiplicities}
\end{table}


\subsection{L\"uscher formula}

An entirely independent approach has been devised by L\"uscher who has
proven an elegant relation between the mass shift of the particle $P$ in a
finite volume $L^3\!\times\!\infty$ and the $P\pi$ scattering amplitude in
infinite volume \cite{Luscher:1985dn} 
\beq
M_P(L)-M_P=-\,{3\ovr16\pi^2\la_P}\;
\int_{-\infty}^\infty\!dy\;\cF_P(\ri y)\,e^{-\sqrt{\Mpi^2+y^2}L}
+O(e^{-\bar ML})
\;.
\label{luscher_mpi_ori}
\eeq
Here $\cF_P(\nu)$ denotes the infinite volume forward ($t\!=\!0$)
scattering amplitude of $P$ and $\pi$ in Minkowski space.
The integration runs along the imaginary axis, i.e.\ $\cF_P(\nu)$ is evaluated
for 
\beq
\nu=\ri y
\label{nuisiy}
\eeq
with real $y$, thus staying far away from the cuts. Only the real part of
$\cF_P(\ri y)$ contributes to the integral, since the imaginary part is odd
in $y$. An additional piece in the original formula, referring to the
3-particle vertex, is omitted here, since we assume $P$ to be a
pseudo-Goldstone boson. The L\"uscher formula (\ref{luscher_mpi_ori}) keeps
only the leading term in an expansion in (fractional) powers of
$e^{-\la_\pi}$. The generic bound $\bar M\!\geq\!\sqrt{3/2}\Mpi$ can be
specified to $\bar M\!=\!\sqrt{2}\Mpi$ in a theory with pseudo-Goldstone
bosons only.

Recently, an analogous ``L\"uscher-type'' formula has been derived for the
finite volume shift of the axial-vector decay constant $F_P$.
It reads \cite{Colangelo:2004xr}
\beq
F_P(L)-F_P=\,+{3\ovr8\pi^2\la_P}\;
\int_{-\infty}^\infty\!dy\;\cN_P(\ri y)\,e^{-\sqrt{\Mpi^2+y^2}L}
+O(e^{-\bar ML})
\label{luscher_fpi_ori}
\eeq
where $\cN_P(\nu)$ is derived from the matrix element $\<\pi\pi|A_\mu|P\>$ via
a subtraction prescription we will specify below.
Like in the mass formula (\ref{luscher_mpi_ori}) the finite volume shift of
$F_P$ is expressed in terms of an infinite-volume amplitude, evaluated in the
unphysical (Euclidean) region, thus far away from the cuts.
Again, only the leading term in an expansion in (fractional) powers of
$e^{-\la_\pi}$ is kept. Note the reverse overall sign, compared to
(\ref{luscher_mpi_ori}), and the fact that the net physical effect is opposite
to this prefactor; in other words $M_P(L)\!>\!M_P$ and $F_P(L)\!<\!F_P$.

To predict the shifts $M_P(L)\!-\!M_P$ and $F_P(L)\!-\!F_P$ in a lattice
calculation with a known box length $L$, and thus to correct the data for this
systematic effect, the formulae (\ref{luscher_mpi_ori},~\ref{luscher_fpi_ori})
must be fed with an explicit representation of the amplitudes $\cF_P(\nu)$
and $\cN_P(\nu)$, respectively.
This is the place where ChPT naturally enters, even if one opts for the
L\"uscher approach.
Using existing knowledge about the relevant amplitude at $n$-loop order, one
gets the leading piece, in the $e^{-\la_\pi}$ expansion, of the finite-volume
shift of $M_P(L), F_P(L)$ to $n\!+\!1$-loop order.
For instance, using the tree-level expressions $\cF_\pi(\nu)\!=\!-\Mpi^2/\Fpi^2$
and $\cN_\pi(\nu)\!=\!-2\Mpi/\Fpi$ in 2-flavor ChPT yields
\bea
\Mpi(L)-\Mpi&=&+\,{3\ovr8\pi^2}{\Mpi^2\ovr\Fpi^2 L}\,K_1(\la_\pi)
+O(e^{-\sqrt{2}\la_\pi})
\\
\Fpi(L)-\Fpi&=&-\,{3\ovr2\pi^2}{\Mpi\ovr\Fpi L}\,K_1(\la_\pi)
+O(e^{-\sqrt{2}\la_\pi})
\eea
in agreement \cite{Gasser:1986vb} with the 1-loop chiral expressions
(\ref{mpi}) and (\ref{fpi}).
Because of the ``elevator''-effect in the loop expansion and since the
associated chiral calculation is in infinite volume, it is much easier to
push to higher chiral orders in the L\"uscher-type setup than with a
straightforward ChPT-in-finite-volume calculation \cite{Colangelo:2002hy}. 
This offers a genuine opportunity to compare several chiral orders and thus
to assess the \emph{chiral convergence behavior}.
Indeed, in \cite{Colangelo:2003hf} the finite volume shift in the pion mass
was evaluated, using ChPT at LO/NLO/NNLO for $\cF_\pi(\nu)$ to get the
asymptotic piece of the full chiral expression at 1/2/3-loop order, and it
was found that for some $(\Mpi,L)$-combinations the chiral series converges
well, if at least the NLO input is included.
Still, one might worry whether the non-asymptotic pieces of order
$O(e^{-\sqrt{2}\la_\pi})$ would prove numerically relevant 
\cite{Colangelo:2002hy}.
In \cite{Colangelo:2003hf} a first attempt was made to discriminate
those regions in the $(\Mpi, L)$-plane where higher orders in the chiral
expansion dominate against those regions where terms omitted in the
L\"uscher approach are more important.
Below, we shall present a resummed version of the L\"uscher-type formulae
(\ref{luscher_mpi_ori}, \ref{luscher_fpi_ori}), where the pieces
$\propto\!e^{-\sqrt{2}\la_\pi}$ and $\propto\!e^{-\sqrt{3}\la_\pi}$ are
included and the terms $O(e^{-(\sqrt{3}+1)/\sqrt{2}\cdot\la_\pi})$ estimated.
On this basis a more precise assessment of the relevance of higher loop
corrections versus higher powers of $e^{-\la_\pi}$ can be made.

Note finally that all L\"uscher-type formulae build on the unitarity of the
theory and thus hold for the full (unquenched) theory.
In the (partially) quenched case it seems indispensable to start in the
framework of subsect.\,2.1, but even then the arguments for using the infinite
volume Lagrangian reside on less solid grounds -- see
Ref.\,\cite{Sharpe:1992ft} for a lucid discussion.


\section{The L\"uscher formula resummed}


In L\"uscher's derivation of the asymptotic formula for the finite volume
correction to particle masses the first step is a proof that the leading
exponential term is given by the sum of all diagrams in which only one
propagator is taken in finite volume.
This class of diagrams yields
\beq
M_P(L)-M_P=
-{1\ovr4M_P}\sum_{\mb{n}\neq\mb{0}}\int\!\!{d^4q\ovr(2\pi)^4}\;
e^{\ri\mb{q}\cdot\mb{n}L}\,G_0(q)\Gamma(\hat{p},q,-\hat{p},-q)+\ldots
\label{eq:mpiL}
\eeq
where $G_0(q)\!=\!1/(\Mpi^2+q^2)$ is the full propagator and
$\Gamma(p_1,p_2,p_3,p_4)$ the four-point vertex function in infinite volume.
L\"uscher then concentrates on the leading exponential contributions (those
with $|\mb{n}|=1$), and shows that, if one disregards terms which are
exponentially suppressed with respect to $\exp(-\la_\pi)$, three of the four
integrations in (\ref{eq:mpiL}) can be performed explicitly and the result
(\ref{luscher_mpi_ori}) is obtained. The same reasoning, however, applies
also to all other terms in the sum in (\ref{eq:mpiL}): 
for each of the terms with $|\mb{n}|>1$, one can obtain its leading
exponential contribution by performing exactly the same
steps that L\"uscher did for the $|\mb{n}|=1$ term and work out
three of the four integrations explicitly. It is easy to keep track of the
vector $\mb{n}$ in doing these manipulations, and to get the resummed
formula 
\beq
M_P(L)-M_P=-\,{1\ovr32\pi^2\la_P}\;\sum_{n=1}^{\infty}{m(n)\ovr\sqrt{n}}
\int_{-\infty}^\infty\!dy\;\cF_P(\ri y)\,e^{-\sqrt{n(\Mpi^2+y^2)}L}
+O(e^{-\bar ML})
\label{eq:Rmpin}
\eeq
where $m(n)$ has been given in Tab.\,\ref{tab:multiplicities} and
$\cF_P(\nu)$ is the $P\pi$ forward scattering amplitude as usual.
The extension which we are proposing is done in the same spirit as the
extension of the domain of integration in (\ref{eq:Rmpin}) to infinity (the
contributions from the region $|y| > \sqrt{\bar M^2\!-\!M_\pi^2}$ are beyond
the accuracy of the formula): being of ``kinematical'' nature, the
extension comes at no cost and may be numerically relevant. In this case,
actually, one even obtains an improvement in the algebraic accuracy of the
formula: we now have $\bar M > \sqrt{3} \Mpi$ and not, as before,
$\bar M\!=\!\sqrt{2}\Mpi$. In other words, the terms $O(e^{-\sqrt{2}\la_\pi})$
and $O(e^{-\sqrt{3}\la_\pi})$ are included now. 
To further clarify the meaning of (\ref{eq:Rmpin}) let us list the two main
classes of exponentially suppressed contributions which are still missing:
\begin{enumerate}
\itemsep-1pt
\item All diagrams which have more than one pion loop in finite volume.
Obviously, these contributions start at the two-loop level in the chiral
expansion.
\item Contributions to the integral (\ref{eq:mpiL}) which are due to
singularities in either the propagator $G_0$ or the vertex function $\Gamma$
which are further away from the real axis than $M_\pi$.
These singularities show up only if one considers the vertex function at one
loop, or the propagator at two-loop accuracy and beyond.
All these contributions appear only if one calculates finite-volume
effects at the two-loop level in the chiral expansion.
\end{enumerate}
In the first class we distinguish those diagrams where different
loops factorize (i.e.\ loops which have no propagator in common) from those
which do not. It is easy to see that the former sub-class yields
corrections of order $e^{-2\la_\pi}$. For two-loop diagrams which do not
factorize (e.g. the two-loop sunset diagram) on may rely on the general
discussion by L\"uscher (cf.\ in particular Eqn.\,(2.49) in
Ref.\,\cite{Luscher:1985dn}) and conclude that the sunset diagram decays
exponentially at large $L$ as $\exp(-N_\mr{\tiny sunset} \la_\pi)$ with 
$N_\mr{\tiny sunset} = (\sqrt{3}\!+\!1)/\sqrt{2} \simeq 1.93$.
For the second class we remark that
the singularities neglected in (\ref{eq:Rmpin}) are due to the exchange of
at least two pions, hence starting at $\nu\!=\!\pm2\Mpi$, and this gives
terms of order $e^{-2\la_\pi}$. We have not tried to prove the statement
that the algebraic accuracy of our formula is given by
\beq
\bar M\!=\!\Mpi(\sqrt{3}\!+\!1)/\sqrt{2}
\label{newbound}
\eeq
beyond the two-loop
level, since it appears to us to be a question of academic interest.

More interesting is the question of how accurate the formula
(\ref{eq:Rmpin}) is numerically. A complete analysis of finite-volume
effects for $M_\pi$ and $F_\pi$ at the two loop level, which is currently
under way \cite{fve2loop}, will clarify this point. The partial results we
have so far indicate that the formula is very accurate.  We have tried to
find an algebraic reason for this, and found out that at the two loop level
all diagrams which do not appear in Eq.~(\ref{eq:Rmpin}) are suppressed
(besides an extra exponential factor) by some power of $1/L$. The numerical
results, however, seem to go beyond what one would expect from such an
argument.

An analogy to the low-temperature expansion seems more suggestive.  In the
effective theory the large volume and the low temperature expansions are in
one-to-one correspondence \cite{Gasser:1987zq}.  In
Ref.\,\cite{Schenk:1993ru} Schenk discussed the propagation of pions
through matter in a state of thermal equilibrium at inverse temperature
$\beta$.  If the temperature is not too high, the hadronic phase mainly
consists of pions, with effects of other excitations such as
$K,\eta,\rho...$ exponentially suppressed.
Due to interactions with pions of the heat bath, the effective pion mass
$\mpi(\beta)$ is given by \cite{Schenk:1993ru}
\beq
\mpi(\beta)-\mpi=-{1\ovr2\mpi}\int\!\!{d^3 q\ovr(2\pi)^32\om_q}\;
n_\mr{B}(\om_q)\;T^{I=0}_{\pi\pi}(s)+O(n_\mr{B}^2)
\label{eq:schenk}
\eeq
with $\om_q=\sqrt{\mpi^2+\mb{q}^2}$ and the density
$n_\mr{B}(x)\!=\!1/(e^{\beta x}-1)$.
The details of the pion kinematics will be discussed in the next section, and
the isospin index refers to the $t$-channel.
One immediately verifies that (\ref{eq:schenk}) agrees with the modified
L\"uscher formula in one dimension to first order in the density $n_\mr{B}$
-- the density factor $n_\mr{B}$ in the integrand is the outcome of the
resummation over $\mb{n}$ in (\ref{eq:mpiL}) if $\mb{n}$ is taken as a
one-dimensional vector.
Schenk has carried the expansion of (\ref{eq:schenk}) one step further and
determined the contributions of order $n_\mr{B}^2$ to the pion mass at finite
temperature: in this extension, effects generated by three-body collisions
are explicitly accounted for. It turned out that these effects are
numerically very small, in line with intuition -- the rescattering of three
pions into three pions is a rare process unless the density is very high.
Although the argument cannot be formulated in the same way for the finite
volume case, we do see that the outcome of the numerical analysis is the same.


\section{Meson masses and decay constants in finite volume}


We start with the resummed L\"uscher formulae for the relative finite-size
shift of pseudoscalar masses and decay constants
\bea
R_{M_P}\equiv
{M_P(L)-M_P \ovr M_P} \!&\!=\!&\! -\frac{\Mpi}{32\pi^2 M_P \la_P}
                 \sum_{n=1}^{\infty} \frac{m(n)}{\sqrt{n}}
                 \int_{-\infty}^{\infty} \! d\til y \,\, \cF_P(\ri \til y) 
                 e^{-\sqrt{n(1 + \til y^2)}\,\la_\pi} 
                 + O(e^{-\bar M L})\qquad
\label{RMP}
\\
R_{F_P}\equiv\,\,\,
{F_P(L)-F_P \ovr F_P} \!&\!=\!&\! +\,\frac{\Mpi}{16\pi^2 F_P \la_P}\,
                 \sum_{n=1}^{\infty} \frac{m(n)}{\sqrt{n}}
                 \int_{-\infty}^{\infty} \! d\til y \, \cN_P(\ri \til y) 
                 e^{-\sqrt{n(1 + \til y^2)}\,\la_\pi} 
                 + O(e^{-\bar M L})\qquad
\label{RFP}
\eea
where all symbols on the r.h.s.\ refer to infinite volume quantities.
The multiplicities $m(n)$ are given in Tab.\,\ref{tab:multiplicities},
$\la_P$ has been defined in (\ref{def_la_p}), and the dimensionless
integration variable (which we will be using from here on) relates to the
previous one via $\til y\!=\!y/\Mpi$. The amplitudes
$\cF_P(\til\nu),\cN_P(\til\nu)$ with 
$\til{\nu}\!=\!\nu/\mpi$ are 
\bea
\label{Eq.:Tp_Np}
\cF_P(\til\nu) 
\!&\!=\!&\! T^{I=0}_{\pi P}(0,-4M_P\nu) \nn
\cN_P(\til\nu) 
\!&\!=\!&\!-\ri\bar A^{I=0}_P(0,-4M_P\nu)
\eea
where 
$T^{I=0}_{\pi P}(t,u\!-\!s)$ 
is the $P\pi$-scattering amplitude with zero
$t$-channel isospin, and 
$\bar{A}^{I=0}_P(t,u\!-\!s)$ 
is the subtracted
amplitude for the decay of the meson $P$ with momentum $p$ into two pions in an
isospin zero state (in the $t$-channel) via an axial current insertion.
The subtraction removes the one-particle reducible contribution and is
defined through
\bea
\bar A^{I=0}_P(t,u-s)&=& \frac{p^\mu}{M_P} (\bar A^{I=0}_P)_\mu \nn
(\bar A^{I=0}_P)_\mu &=& (A_P^{I=0})_\mu-\ri Q_\mu F_P 
                         \frac{T^{I=0}_{\pi P}(t,u-s)}{Q^2-M_P^2} \nn
(A_P^{I=0})_\mu      &=& \<(\pi(p_1) \pi(p_2) )_{I=0}|A_\mu(0)|P(p)\>
\label{Eq.:sub}
\eea
with $Q\!=\!p\!-\!p_1\!-\!p_2$. Notice that the axial current in
(\ref{Eq.:sub}) must be normalized such that $\langle 0 | A_\mu(0) | P(p)
\rangle= \ri p_\mu F_P$. The amplitudes
$\cF_P(\til\nu),\cN_P(\til\nu)$ for $P=\pi,K,\eta$ have all (with the
exception of $\cN_\eta(\til\nu)$, see below) been calculated at least at
the one-loop level in ChPT. They have the generic form
\begin{eqnarray}
\cF_P(\til\nu) &=& \;\!\cF_P^{(2)}(\til{\nu}) +
                 \;\!\xi_P \cF_P^{(4)}(\til{\nu}) +
                 \;\!\xi_P^2 \cF_P^{(6)}(\til{\nu}) +
                 \mathcal{O}(\xi_P^3) \nn
\cN_P(\til\nu) &=& \cN_P^{(2)}(\til{\nu}) +
                 \xi_P \cN_P^{(4)}(\til{\nu}) +
                 \xi_P^2 \cN_P^{(6)}(\til{\nu}) +
                 \mathcal{O}(\xi_P^3) \nonumber
\end{eqnarray}
with $\xi_P$ defined in (\ref{def_xi_p}).
Inserting such an expansion of the amplitude in (\ref{RMP}, \ref{RFP}) leads to
\begin{eqnarray}
  R_{M_P} &=& -\sum_{n=1}^{\infty}
  \frac{m(n)}{2\sqrt{n}}\frac{1}{\lambda_\pi}
  \frac{\mpi}{M_P}  \xi_P  \Big[I_{M_P}^{(2)} 
  + \xi_P I_{M_P}^{(4)} +\xi_P^2 I^{(6)}_{M_P}+\ordo (\xi_P^3)\Big]
\label{generic_M}
\\
  R_{F_P} &=& +\sum_{n=1}^{\infty}
  \frac{m(n)}{\sqrt{n}}\frac{1}{\lambda_\pi}\,
  \frac{\fpi}{F_P}\,\xi_\pi\,\Big[\,I_{F_P}^{(2)}
  + \,\xi_P I_{F_P}^{(4)}+ \,\xi_P^2 I^{(6)}_{F_P} \,+ \ordo (\xi_P^3)\Big]
\label{generic_F}
\end{eqnarray}
where the $I_{M_P}^{(2/4/6)},I_{F_P}^{(2/4/6)}$ can be written in terms of a
few basic integrals, as reported in Sect.\,\ref{sec:analytic}.
Please note the relative factor $2$ in these two equations, to be consistent
with \cite{Colangelo:2003hf,Colangelo:2004xr}.
In the following we elaborate on the explicit form of (\ref{Eq.:Tp_Np}) for
the pion, kaon and eta.


\subsection{Pion}

The amplitudes $\cF_\pi(\til\nu),\cN_\pi(\til\nu)$ defined in
Eq.\,(\ref{Eq.:Tp_Np}) have been given in
Refs.\,\cite{Colangelo:2003hf,Colangelo:2004xr}, respectively.
We give them here for convenience. 

\subsubsection{Pion mass}

Consider (Minkowski space) $\pi\pi$-scattering
\begin{equation}
\pi(p_1)+\pi(p_2) \rightarrow \pi(p_3) + \pi(p_4) \nn
\end{equation}
with the kinematics
\beq
s = (p_1+p_2)^2 \; , \qquad t = (p_1-p_3)^2 \; , \qquad u = (p_1-p_4)^2 \fs
\label{def_mandelstam}
\eeq
Isospin decomposition allows one to construct the $t$-channel isospin zero
amplitude
\begin{equation}
T^{I=0}_{\pi\pi}(t,u-s) = A(s,t,u) + 3A(t,s,u) + A(u,s,t)
\end{equation}
from the invariant amplitude $A(s,t,u)$ \cite{Bijnens:pipi}.
The amplitude entering the L\"uscher formula follows by imposing the forward
scattering kinematics $t\!=\!0$, viz.\
\begin{equation}
\cF_\pi(\til\nu) = T^{I=0}_{\pi\pi}(0,-4\mpi \nu) \fs
\label{Tpi}
\end{equation}

\subsubsection{Pion decay constant}

We adopt the notation of Ref.\,\cite{Colangelo:2004xr}, which relates to the
one used in Eq.\,(\ref{Eq.:sub}) by means of the one pion in the initial state
being transferred to an outgoing pion.
Crossing symmetry relates the two via $p_3\!=\!-p$.
The amplitude for the creation of three pions out of the vacuum with an axial
current has been performed up to NLO in Ref.\,\cite{Colangelo:1996hs}.
It is decomposed according to
\beq
\< \pi^1(p_1) \pi^1(p_2) \pi^3(p_3) | A^3_\mu(0) | 0 \> =
(p_1+p_2)_\mu\,G + (p_1-p_2)_\mu\,H + p_{3 \mu}\,F
\label{eq:A3pi}
\eeq
with the three scalar functions
\footnote{Note that our functions $F,G,H$ are $1/\sqrt{2}$ times those of
Ref.\,\cite{Colangelo:1996hs}, in agreement with the notation in
Ref.\,\cite{Colangelo:2004xr}.}
$F\!=\!F(s_1,s_2,s_3)$, $G\!=\!G(s_1,s_2,s_3)$ and $H\!=\!H(s_1,s_2,s_3)$.
The superscripts on the pion states and axial current are isospin indices.
We have employed the variables $s_1\!=\!(p_2+p_3)^2$ and cyclic
permutations. The combination which has two of the outcoming pions in an
$I\!=\!0$ state in the $s_3$-channel 
is given by
\bdm
\<(\pi(p_1)\pi(p_2))_{I=0}\pi^3(p_3)|A^3_\mu(0)|0\> = (A_\pi^{I=0})_\mu
\edm
\beq
(A_\pi^{I=0})_\mu =
(p_1+p_2)_\mu\,G_0(s_1,s_2,s_3)+
(p_1-p_2)_\mu\,H_0(s_1,s_2,s_3)+
(p_3)_\mu    \,F_0(s_1,s_2,s_3)
\eeq
with the isospin projected
\footnote{The relation for $G_0$ is given for completeness; on imposing forward
scattering kinematics, it will drop out.}
form factors
\bea
F_0(s_1,s_2,s_3)\!&\!=\!&\! 3F_{123}+G_{231}+G_{312}-H_{231}+H_{312}
\nn
G_0(s_1,s_2,s_3)\!&\!=\!&\! 3G_{123}+\frac{1}{2}
\big[ F_{231}+F_{312}+G_{231}+G_{312}+H_{231}-H_{312} \big]
\nn
H_0(s_1,s_2,s_3)\!&\!=\!&\! 3H_{123}+\frac{1}{2}
\big[ F_{231}-F_{312}-G_{231}+G_{312}-H_{231}-H_{312} \big]
\nonumber
\eea
where $X_{ijk}\!=\!X(s_i,s_j,s_k)$ and $X\!=\!F,G,H$.
The pole that the amplitude $(A_\pi^{I=0})_\mu$ has in the unphysical region
$Q^2\!=\!\mpi^2$, $Q=-(p_1+p_2+p_3)$, needs to be subtracted as specified
in (\ref{Eq.:sub})
\beq
(\bar{A}_\pi^{I=0})_\mu = (A_\pi^{I=0})_\mu - \ri Q_\mu\fpi
\frac{T_{\pi\pi}^{I=0}(s_3,s_1-s_2)}{Q^2 - \mpi^2 }
\label{Eq:subFp}
\eeq
and in the end the result is evaluated in the forward kinematic configuration,
i.e.\ for $s_3\!=\!0$.
Hence the function $\bar{A}_\pi^{I=0}$ is a function of just one variable
$\nu\!=\!(s_2-s_1)/(4\mpi)$, viz.\
\bea
\bar{A}_\pi^{I=0}(0,-4\mpi\nu) &=& 2 \nu h_0(\nu) + \mpi \bar{f}_0(\nu)
\\
h_0(\nu)               &=&  H_0(2\mpi(\mpi-\nu),2\mpi(\mpi+\nu),0)
\nn 
\bar f_0(\nu)          &=&  \,\bar F_0(2\mpi(\mpi-\nu),2\mpi(\mpi+\nu),0)
\nonumber
\eea
where the bar on the $F_0$ form factor indicates that it is defined after the
subtraction of the pion pole ($H_0$ remains unaffected).
The amplitude which enters the formula for $F_\pi(L)$ is then
\beq
\cN_\pi(\til\nu)=-\ri\,\bar{A}_\pi^{I=0}(0,-4\mpi\nu)
\;.
\eeq


\subsection{Kaon}

The amplitudes $\cF_K(\til\nu)$ and $\cN_K(\til\nu)$ may be extracted from
the form factors of the $K_{l4}$ decay, as calculated in
Ref.\,\cite{Bijnens:1994ie} up to NLO.  Furthermore, an explicit
representation of $\cF_K(\til\nu)$ is found in the $\pi K$-scattering study
of Ref.\,\cite{Bernard:1990kw} which has been extended to NNLO in
Ref.\,\cite{Bijnens:2004bu}.  Below, we stick to the NLO input, both for
$\cF_K(\til\nu)$ and $\cN_K(\til\nu)$.  For the mass the finite volume
corrections at 2-loop level are very small (see discussion below) and a
NNLO input is therefore not of particular interest for a lattice
application.  For the decay constant the result of the L\"uscher formula
with NLO input is not particularly small (see below), and a NNLO refinement
would be useful.  However, in this case one of the ingredients (to be
precise: one of the 2-loop form factors of the axial-vector matrix element
discussed in \cite{Amoros:2000mc}) is missing.  In the following we
establish the relation of the amplitudes $\cF_K(\til\nu)$ and
$\cN_K(\til\nu)$ to results given in the literature.

\subsubsection{Kaon mass}

The $\pi K$-scattering amplitude $T_{\pi K}(s,t,u)$ has been calculated at
NLO in Ref.\,\cite{Bernard:1990kw} \footnote{As noted in the literature,
there are two typos in (3.16) of \cite{Bernard:1990kw}: The prefactor of
$(M_{\pi K}^r(u)\!-\!M_{K\et}^r(u))$ should read $(\mk^2\!-\!\mpi^2)^2$,
and the term multiplying ${3\ovr8}J_{K\et}^r(u)$ is
$(u\!-\!{2\ovr3}(\mpi^2\!+\!\mk^2))^2$ [the latter is correct in the
preprint].}
\begin{equation}
 \pi^+(p_1) + K^+(p_2) \rightarrow \pi^+(p_3) + K^+(p_4)
\end{equation}
with the Mandelstam variables (\ref{def_mandelstam}).
This gives the $t$-channel isospin zero amplitude via
\begin{equation}
T^{I=0}_{\pi K}(t,u-s) = \frac{3}{2} [ T_{\pi K}(s,t,u) + T_{\pi K}(u,t,s) ]
\end{equation}
from which the amplitude entering the L\"uscher formula follows by applying
the forward scattering kinematics, $t = 0$, viz.\
\begin{equation}
\cF_K(\til\nu) = T^{I=0}_{\pi K}(0,-4\mk \nu) \fs 
\end{equation}

\subsubsection{Kaon decay constant}

We need the matrix element of an axial current between a kaon and two pions,
which occurs in the evaluation of the $K_{l4}$ decay.
Ref.\,\cite{Bijnens:1994ie} defines the three scalar form factors $F,G,R$
through
\bdm
(A_K)_\mu= \frac{1}{\sqrt{2}}
\< \pi^+(p_1)\pi^-(p_2)|A_\mu^{4-\ri5}(0)|K^+(p)\>
\edm
\beq
(A_K)_\mu={-\ri\ovr \sqrt{2} \mk}
\Big[ (p_1+p_2)_\mu\,F + (p_1-p_2)_\mu\,G + Q_\mu\,R \Big]
\eeq
with $F\!=\!F(s,t,u), G\!=\!G(s,t,u), R\!=\!R(s,t,u)$ and
the kinematic variables
\beq
s = (p_1+p_2)^2 \;,\qquad t = (p_1-p)^2 \;,\qquad  u = (p_2-p)^2 \;,
\qquad Q = p-p_1-p_2 \fs
\eeq
The combination with the pions in an $s$-channel 
isospin zero state is
\beq
(A_K^{I=0})_{\mu}= {\ri\ovr \sqrt{2} \mk}
\Big[ (p_1+p_2)_\mu\,F^+ + (p_1-p_2)_\mu G^- +Q_\mu R^+ \Big]
\eeq
where
\bdm
X^{\pm} = {1\ovr2}[X(s,t,u)\pm X(s,u,t)]
\edm
with $X=F,G,R$.
Subtracting the pole at $Q^2=(p-p_1-p_2)^2=\mk^2$ as defined in (\ref{Eq.:sub})
and evaluating the amplitude $\bar{A}^{I=0}_K(s,t-u)$ in the forward
scattering configuration $s=0$ yields
\begin{equation}
\bar{A}^{I=0}_K(0,-4\mk\nu) = 
-{3\ri\ovr \sqrt{2} \mk} \Big[ 2\nu g^-(\nu) + \mk \bar{r}^+(\nu) \Big]
\end{equation}
with
\begin{equation}
g^-(\nu) = G^-(0,\mpi^2+\mk^2-2\mk\nu,\mpi^2+\mk^2+2\mk\nu)
\end{equation}
and analogously for $\bar{r}^+$. Here, the bar on the form factor $R^+$
indicates again that it is defined after subtraction of the kaon pole (the
form factor $G^-$ remains unaffected by the subtraction). Finally,
according to Eq.(\ref{Eq.:Tp_Np}), the amplitude entering the L\"uscher
formula is 
\beq
\cN_K(\til\nu)=-\ri\bar A_K^{I=0}(0,-4\mk\nu)
\;.
\eeq


\subsection{Eta}

The decay constant of the $\eta$ is of no phenomenological interest, like
for the other neutral members of the pseudoscalar octet. We therefore
refrain from discussing the finite volume effects for this quantity (as a
side remark, we notice that also the analogue of the $K_{l4}$ decay
amplitude, the $\langle (2 \pi)_{I=0} | A_\mu^8 | \eta \rangle $ amplitude,
is of no phenomenological interest and has never been calculated).
We restrict ourselves to the finite volume effects on the $\eta$ mass.

The $\pi\eta$-scattering amplitude $T_{\pi\eta}(s,t,u)$ has been calculated to
NLO in Ref.\,\cite{Bernard:1991xb},
\begin{equation}
\pi^0(p_1) + \eta(p_2) \rightarrow \pi^0(p_3) + \eta(p_4) \nn
\end{equation}
with kinematic variables (\ref{def_mandelstam}),
and relates to the isospin zero amplitude in the $t$-channel through
\begin{equation}
T_{\pi\eta}^{I=0}(t,u-s) = 3 T_{\pi\eta}(s,t,u) \fs
\end{equation}
The amplitude entering the L\"uscher formula follows by applying the
forward scattering kinematics, $t = 0$, viz.\
\begin{equation}
\cF_\eta(\til\nu) = T^{I=0}_{\pi\eta}(0,-4M_\eta \nu) \fs 
\end{equation}


\section{Summary of the analytical results}\label{sec:analytic}


In order to use the asymptotic formulae we have to feed them with the
specific expressions for the scattering amplitude or the axial vector matrix
element that are available in the literature. In this section we present
such explicit formulae for $M_P(L)$ and $F_P(L)$ in two versions. We start
with the complete expressions with some of the lengthier parts
relegated to the appendix. The second step entails simplified versions
of the unhandy parts, together with a discussion of how they relate to the
complete version.


\subsection{Full formulae}

Evaluating (\ref{Tpi}) in the $SU(2)$ framework the fractional shift of the
pion mass takes the form (\ref{generic_M}) with
\bea
I^{(2)}_{M_\pi}&=&-B^0
\nn
I^{(4)}_{M_\pi}&=&B^0
\bigg[
-{55\ovr18}+4\lb_1+{8\ovr3}\lb_2-{5\ovr2}\lb_3-2\lb_4
\bigg]
+B^2
\bigg[
{112\ovr9}-{8\ovr3}\lb_1-{32\ovr3}\lb_2
\bigg]
+S^{(4)}_{M_\pi}
\nn
I^{(6)}_{M_\pi}&=&B^0
\bigg[
{10049\ovr1296}-{13\ovr72}N
+{20\ovr9}\lb_1-{40\ovr27}\lb_2-{3\ovr4}\lb_3
-{110\ovr9}\lb_4-{5\ovr2}\lb_3^{\,2}-5\lb_4^{\,2}\nn
&&
+\Big(16\lb_1+{32\ovr3}\lb_2-11\lb_3\Big)\lb_4
+\ell_\pi
\Big({70\ovr9}\ell_\pi+12\lb_1+{32\ovr9}\lb_2-\lb_3+\lb_4+{47\ovr18}\Big)\nn
&&
+5\rt_1+4\rt_2+8\rt_3+8\rt_4+16\rt_5+16\rt_6\bigg]\nn
&&
+B^2
\bigg[
{3476\ovr81}-{77\ovr288}N
+{32\ovr9}\lb_1+{464\ovr27}\lb_2+{448\ovr9}\lb_4
-{32\ovr3}(\lb_1+4\lb_2)\lb_4\nn
&&
+\ell_\pi
\Big({100\ovr9}\ell_\pi+{8\ovr3}\lb_1+{176\ovr9}\lb_2-{248\ovr9}\Big)
-8\rt_3-56\rt_4-48\rt_5+16\rt_6
\bigg]
+S^{(6)}_{M_\pi}
\label{explicit_Mpi}
\eea
where we use the abbreviations ($K_i$ denotes the modified Bessel function)
\bea
I_{M_P}^{2/4/6}&\equiv&I_{M_P}^{2/4/6}(\sqrt{n}\la_\pi)\;,\qquad
B^{0/2}\equiv B^{0/2}(\sqrt{n}\la_\pi)
\label{abbrev1}
\\
B^0(x)&=&2K_1(x)\;,\qquad\qquad\hspace{-1pt}
B^2(x)=2K_2(x)/x
\\
\ell_P&=&\ln(M_P^2/\mu^2)\;,\qquad\qquad
N=16\pi^2
\label{abbrev2}
\eea
and the $\bar\ell_i$ which carry a mild logarithmic quark mass dependence
\cite{Colangelo:2001df}
\beq
\bar\ell_i=\bar\ell_i^\mr{phys}+2\log\Big({\Mpi^\mr{phys}\ovr\Mpi}\Big)
\label{barelli}
\eeq
with mass independent $\bar\ell_i^\mr{phys}$ as given in
Tab.\,\ref{tab:SU2SU3}.
The terms $S^{(4)}_{M_\pi},S^{(6)}_{M_\pi}$ are contributions from the loop
functions at order $p^4,p^6$.
They are explicitly given in appendix \ref{app:In} and their numerical
importance is discussed below.
Formula (\ref{explicit_Mpi}) has already appeared in
\cite{Colangelo:2003hf}.

The fractional shift in the pion decay constant takes the form
(\ref{generic_F}) with
\bea
I^{(2)}_{F_\pi}&=&-2 B^0 \nn
I^{(4)}_{F_\pi}&=&B^0\left[-\frac{7}{9} + 2 \lb_1+ \frac{4}{3}
\lb_2 - 3 \lb_4 \right]
+ B^2 \left[ \frac{112}{9} - \frac{8}{3} \lb_1-
\frac{32}{3} \lb_2 \right] + S^{(4)}_{F_\pi}
\label{explicit_Fpi}
\eea
and $S^{(4)}_{F_\pi}$ moved to the appendix.
Formula (\ref{explicit_Fpi}) has already been given in
\cite{Colangelo:2004xr}. 

With the abbreviation $x_{PQ}=M_P^2/M_Q^2$ the finite volume shift of the kaon
mass is given by
\bea
  I_{\mk}^{(2)} &=& 0\nn
  I_{\mk}^{(4)} &=& 3 x_{\pi K}^{1/2} \bigg\{
   B^0 \bigg[  \frac{x_{\pi K}}{9}
  + 8 N x_{\pi K}
  (4 L_1^\mr{r}+L_3^\mr{r}-4L_4^\mr{r}-L_5^\mr{r}+4L_6^\mr{r}+2L_8^\mr{r})\nn
&&+ \ell_\pi \frac{x_{\pi K}^2}{4 (1-x_{\pi K})} + \frac{\ell_K}{16} 
\Big(-\frac{4}{1-x_{\pi K}} + \frac{1-10 x_{\pi K}+x_{\pi K}^2}{6(x_{\eta
      K}-1)} + \frac{7+x_{\pi K}}{2}  \Big)\nn
&&+ \frac{\ell_\eta}{32} \Big(\frac{2}{3}+(1-x_{\pi K}) (x_{\eta K}-1)
  +\frac{53}{9} x_{\pi K} -\frac{x_{\pi K}^2}{3}
 -\frac{1-10 x_{\pi K}+x_{\pi K}^2}{3 (x_{\eta K}-1)} \Big)
                              \bigg]\nn
&&+ B^2 \bigg[
                - 8 N x_{\pi K}  (4 L_2^\mr{r}+L_3^\mr{r})  
                -\ell_\pi \frac{5 x_{\pi K}^2}{2 (1-x_{\pi K})}\nn
                &&+ \ell_K \frac{x_{\pi K}}{2} \Big(\frac{5}{1-x_{\pi K}}
                -\frac{1}{x_{\eta K}-1} \Big)
                +  \ell_\eta \frac{x_{\pi K} x_{\eta K}}{2 (x_{\eta K}-1)}
                \bigg] \bigg\} + S^{(4)}_{M_K}
\label{explicit_Mka}
\eea
and the finite volume correction for $F_K$ takes the form (\ref{generic_F})
with
\bea
  I_{F_K}^{(2)} &=& -\frac{3}{4}B^0  \\
  I_{F_K}^{(4)} &=& \! \! \! 
  B^0 \bigg[ \frac{3}{16} x_{\pi K} 
  \Big( \frac{\ell_K- x_{\pi K} \ell_\pi}{1- x_{\pi K}}
  +\frac{\ell_\eta-x_{K \eta} \ell_K}{1-x_{K \eta}}  + 2
  \ell_\pi \Big( x_{ \pi \eta} - 
    \frac{9}{4} \Big) \Big)  \nn
  &+& \! \! \! \frac{3}{32} (2 \ell_K + 3 x_{\eta K} \ell_\eta) 
  + 12 N x_{\pi K} (4 L_1^\mr{r} + L_3^\mr{r} -2 L_4^\mr{r} )
  -3N L_5^\mr{r} (1+x_{\pi K})  \bigg] \nn
  &+&  \! \! \!  B^2 x_{\pi K} 
  \bigg[\frac{15}{2} \frac{\ell_K-x_{\pi K} \ell_\pi}{1-x_{\pi K}}
  +\frac{3}{2}\frac{ \ell_\eta- x_{K \eta}\ell_K}{1-x_{K \eta}}
  - 24 N (4 L_2^\mr{r}+ L_3^\mr{r} ) \bigg] + S^{(4)}_{F_K}
\label{explicit_Fk}
\eea
and $S^{(4)}_{M_K}, S^{(4)}_{F_K}$ given in App.\,\ref{app:In} and
the convention (\ref{abbrev1}) applied throughout, i.e.\
$I_{X_K}^{(4)}\!\equiv\!I_{X_K}^{(4)}(\sqrt{n}\la_\pi)$ and
$S_{X_K}^{(4)}\!\equiv\!S_{X_K}^{(4)}(\sqrt{n}\la_\pi)$.
Analogously, the eta mass in finite volume is given through
\bea
  I_{\me}^{(2)} &=&  x^{3/2}_{\pi\eta} B^0\nn
  I_{\me}^{(4)} &=&  x_{\pi\eta}^{3/2} \bigg\{
  B^0 \bigg[-\frac{2+x_{\pi\eta}}{3}
         +  \ell_\pi x_{\pi\eta} \Big(\frac{2}{3(1-x_{\pi\eta})}
         -\frac{13}{6}\Big)  \nn
&&       +  \ell_K (2 x_{K\eta}-x_{\pi\eta}) +  \ell_\eta
         \Big(\frac{x_{\pi\eta}}{6}-\frac{2}{3(1-x_{\pi\eta})}\Big) \nn
&&       +  16 N \Big(6 (L_1^\mr{r}-L_4^\mr{r}+L_6^\mr{r}-L_7^\mr{r})
                      +L_3^\mr{r}-L_5^\mr{r}
                           +x_{\pi\eta} (6 L_7^\mr{r}+3 L_8^\mr{r})\Big)
         \bigg]\nn
&&+ B^2 \bigg[ 9 (1+\ell_K)
         -  32N (3 L_2^\mr{r}+L_3^\mr{r})
         \bigg] \bigg\} + S^{(4)}_{M_\eta} \fs
\label{explicit_Met}
\eea
Comparing (\ref{explicit_Mka}) and (\ref{explicit_Met}) to (\ref{explicit_Mpi})
it is obvious that the $SU(3)$ breaking renders the expressions for
$\mk(L),\me(L)$ substantially more complicated than for $\mpi(L)$.
It is remarkable that to leading order the finite volume correction for
the kaon mass vanishes [in the theory with virtual pions only, cf.
Eq.~(\ref{RMka}) in app.~\ref{app:KE}], while for the eta mass there is a
suppression factor $(M_\pi/M_\eta)^2$ [cf. Eq.~(\ref{RMet}) in
app.~\ref{app:KE}].
As we shall see in the numerical analysis, these finite volume corrections are
practically negligible.


\subsection{Simplified formulae}\label{sec:simplified}

Beyond tree-level, the chiral representation of the amplitudes
$\cF_P(\til\nu), \cN_P(\til\nu)$ that enter the resummed asymptotic
formulae tend to become rather complicated.  As a result, the expressions
$S^{(4/6)}_{M_P}$ and $S^{(4)}_{F_P}$ are not particularly handy, see
App.~\ref{app:In}.  In the immediate vicinity of $\nu\!=\!0$, however, a
polynomial approximation to the chiral amplitudes reproduces them rather
well.  The reason behind is that the nonanalytic structure closest to the
origin is the cut starting at $\nu\!=\!\pm\Mpi$. Therefore, for imaginary
$\nu$ close to the origin [i.e.\ for $y\!\in\![-\Mpi,\Mpi]$ with $y$
relating to $\nu$ via (\ref{nuisiy})] even a second order polynomial
reproduces the amplitude rather accurately, while outside this region the
quality of the representation does not matter, due to the suppression
factor $\exp(-\sqrt{n(\Mpi^2+y^2)}L)$.  The advantage of such a polynomial
representation is that all integrals can be performed analytically, and
there is an analytic bound on the remainder.

For the pion mass and decay constant we find
\bea
S^{(4)}_{M_\pi}&=&\frac{13}{3}g_0B^0-\frac{1}{3}
\Big(40g_0+32g_1+26g_2\Big)B^2+O\Big(B^4(\sqrt{n}\la_\pi)\Big)
\nn
S^{(4)}_{F_\pi}&=&\frac{1}{6}\Big(8g_0-13g_1\Big)B^0
-\frac{1}{3}\Big(40g_0-12g_1-8g_2-13g_3\Big)B^2
+O\Big(B^4(\sqrt{n}\la_\pi)\Big)
\eea
where the coefficients $g_i$ are the Taylor coefficients of the function
\footnote{The function $g(x)$ relates to the standard scalar one-loop
function $\bar J$ through $g(x)\!=\!16\pi^2\bar J(x\mpi^2)$.} 
\beq
g(x)=\sigma \log \frac{\sigma -1}{\sigma+1} + 2
\label{def_g_gprime}
\eeq
with $\sigma=\sqrt{1-4/x}$ around the point $x=2$
\be
g(2+\epsilon )=g_0+ g_1 \epsilon + {1\ovr2}g_2 \epsilon^2
+{1\ovr6}g_3 \epsilon^3 +O(\epsilon^4)
\ee
with the explicit values
\be
g_0=2-{\pi\ovr2}\;,\quad
g_1={\pi\ovr4}-{1\ovr2}\;,\quad
g_2={1\ovr2}-{\pi\ovr8}\;,\quad
g_3={3\pi\ovr16}-{1\ovr2}
\;.
\ee
For $S^{(6)}_{M_\pi}$ a similar short-hand version follows 
in the same manner. We refrain from showing them here, because these
contributions turn out to be numerically so small that one could simply drop
them. In
$S_{\mk}^{(4)},S_{\fk}^{(4)}$ and $S_{\me}^{(4)}$ a polynomial expansion would
not really simplify the representation, due to the
different meson masses involved in the loop functions. 


\section{Numerical analysis}


We are now in a position to evaluate the formulae for the relative finite
volume corrections to $M_P$ and $F_P$, as presented in the previous
section.  To fully specify the meaning of our formulae we need to give, as
the last ingredient, the quark mass dependence of the infinite-volume
quantities $\Mpi,\mk,\me,\fpi,\fk$.  In line with the setup of our
calculation we use 2-flavor ChPT for the pion mass and decay constant and
3-flavor ChPT for the kaon and eta counterparts.  In either case the
low energy parameters are determined from phenomenology and summarized in
Tab.\,\ref{tab:SU2SU3}.  Regarding the $SU(2)$ low energy constants
$\bar{\ell}_i$ and $\til{r}_i(\mu)$ we use the values obtained in
Ref.\,\cite{Colangelo:2001df,Bijnens:pipi}; for the $SU(3)$ low energy
constants $L_i^\mr{r}(\mu)$ we refer to the $O(p^4)$ fit of
Ref.\,\cite{Amoros:2001cp}.  In the former case, the full correlation
matrix is given in \cite{Colangelo:2001df} in the latter case it has been
communicated privately \cite{BijnensPrivate}, but in either case the final
errors are almost the same, regardless whether the full correlation matrix
is used or just the diagonal part.

\begin{table}
\begin{center}
\begin{tabular}{|c|r|}
\hline
$i$ & \phantom{$\bar{\bar{\bar{X}}}\!\!$}$\bar{\ell}_i^\mr{phys}$\\[1mm]
\hline
1&$-0.4\pm0.6$\\
2&$ 4.3\pm0.1$\\
3&$ 2.9\pm2.4$\\
4&$ 4.4\pm0.2$\\
\hline
\end{tabular}
\hspace{1cm}
\begin{tabular}{|c|r|}
\hline
$i$ & \phantom{$\bar{\bar{\bar{X}}}\!\!$}$\til{r}_i(M_\rho)$\\[1mm]
\hline
1&$-1.5\times(1\pm1)$\\
2&$ 3.2\times(1\pm1)$\\
3&$-4.2\times(1\pm1)$\\
4&$-2.5\times(1\pm1)$\\
5&$ 3.8\pm1.0$\\
6&$ 1.0\pm0.1$\\
\hline
\end{tabular}
\hspace{1cm}
\begin{tabular}{|c|r|}
\hline
$i$ & \phantom{$\bar{\bar{\bar{X}}}\!\!$}$L_i^\mr{r}(M_\rho)\cdot 10^3$\\[1mm]
\hline
1&$ 0.38\pm0.18$\\
2&$ 1.59\pm0.15$\\
3&$-2.91\pm0.32$\\
4&$ 0.00\pm0.80$\\
5&$ 1.46\pm0.10$\\
6&$ 0.00\pm0.30$\\
7&$-0.49\pm0.24$\\
8&$ 1.00\pm0.21$\\
9&$ 6.90\pm0.70$\\
\hline
\end{tabular}
\end{center}
\vspace{-4mm}
\caption{$SU(2)$-framework: values at the physical pion mass of the low energy
constants $\bar{\ell}_i$ from \cite{Colangelo:2001df} together with
the $p^6$ low energy constants $\til r_i(\mu\!=\!M_\rho)$.
Note that the $\bar{\ell}_i$ used in the formulae for $R_{\mpi},R_{\fpi}$
differ from these values by a term logarithmic in $\mpi^{}/\mpi^\mr{phys}$,
see (\ref{barelli}).
$SU(3)$-framework: $L_i^\mr{r}(\mu\!=\!M_\rho)$ taken from the $O(p^4)$
fit in \cite{Amoros:2001cp} (the uncertainties in $L_6$ and $L_9$ are our
estimate).}
\label{tab:SU2SU3}
\end{table}


\subsection{$\mpi$ dependence of $\fpi$}

The quark mass dependence of $\mpi$ and $\fpi$ has been computed, up to 2
loops, in Refs.\,\cite{Gasser:1983yg,Burgi:1996qi,Colangelo:2001df}.
What we need in the present context is $\fpi$ as a function of $\mpi$, i.e.\
the single relationship that one gets after eliminating the quark mass.
This relation has been given in \cite{Colangelo:2003hf}, where also the
$SU(2)$ low energy constant $F\!=\!(86.2\pm0.5)\,\MeV$ has been found.
We stress that with $\mpi$ (or $\fpi$) we mean simultaneously the pion mass
(decay constant) in an infinite volume lattice simulation and in ChPT to the
highest loop order available, i.e.\ to $O(p^6)$ in the $SU(2)$ framework.
Finally, we mention that the chiral expansion parameter
$\xi_\pi \!\equiv\!(\mpi/4\pi\fpi)^2$ remains small for pion masses up to
$500\MeV$, see the discussion in \cite{Colangelo:2003hf} for details.


\subsection{$\mpi$ and $m_s$ dependence of $\mk$, $F_K$ and $\me$}

\begin{figure}
\centering
\includegraphics[width=8.3cm]{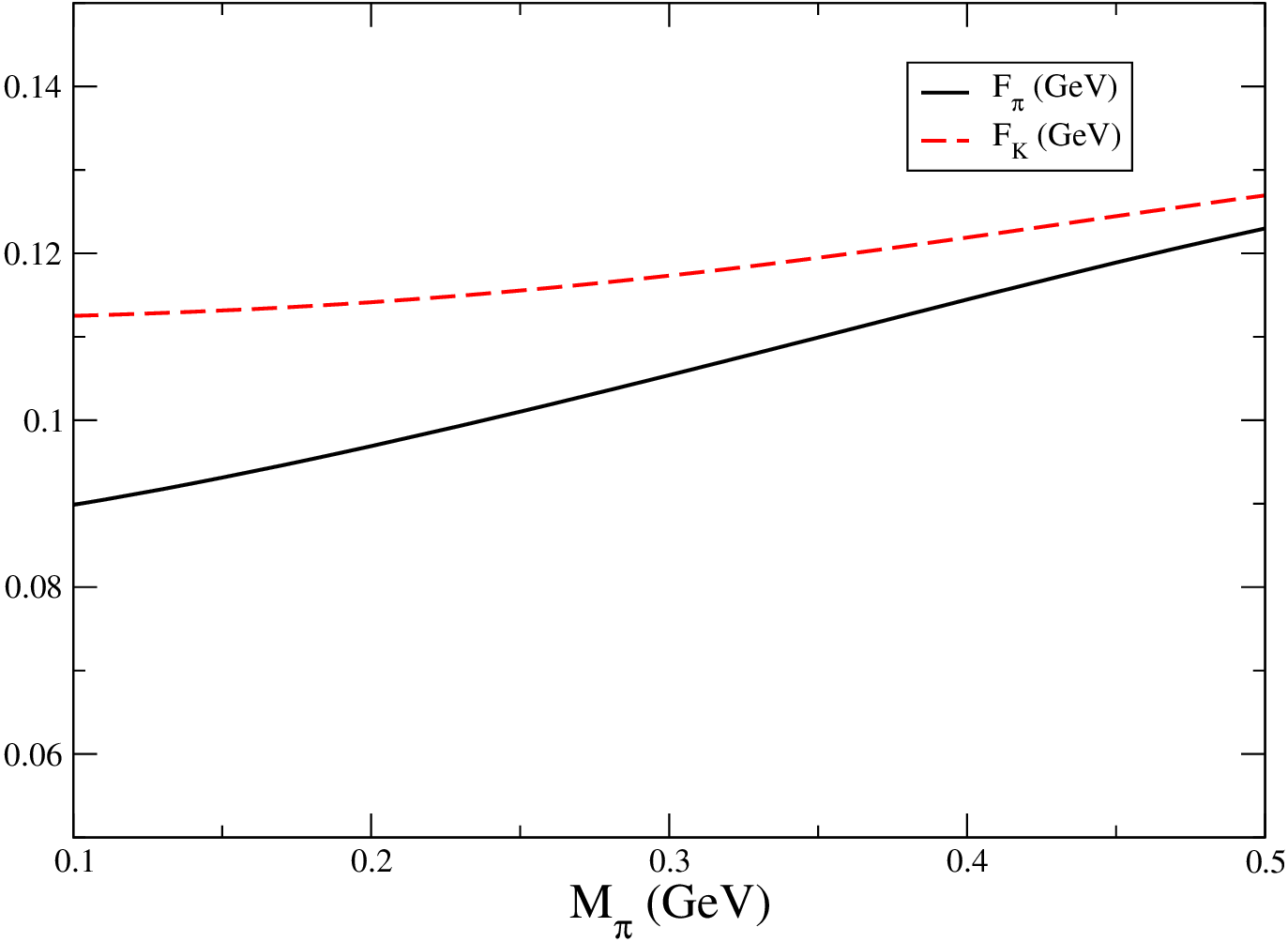}
\includegraphics[width=8.3cm]{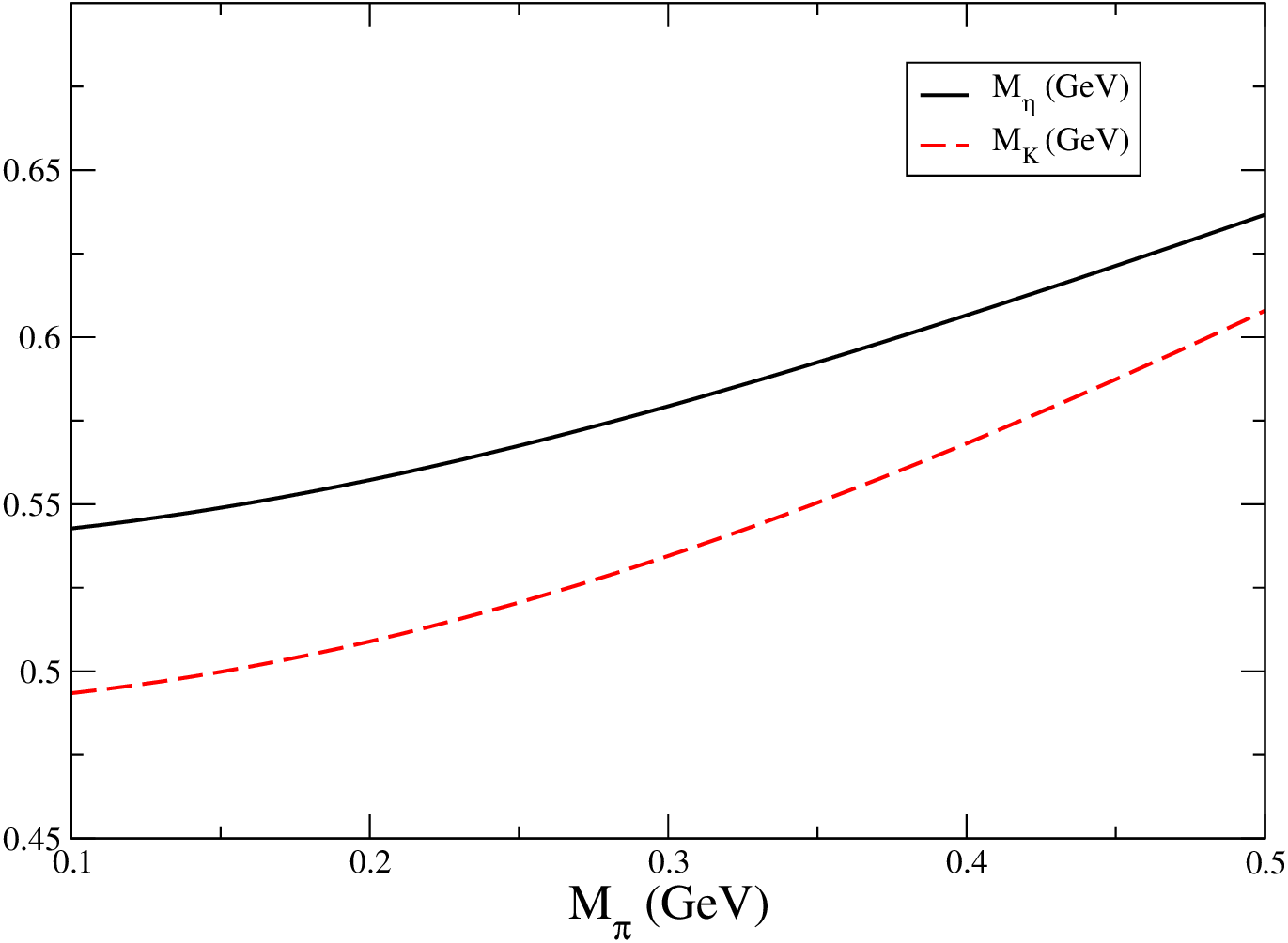}
\vspace{-2mm}
\caption{$\mpi$ dependence (in infinite volume) of $\fpi,\fk,\mk,\me$. For the
latter three quantities the strange quark mass has been fixed as to reproduce
$\mk^\mr{phys}$ at $\mpi^{}\!=\!\mpi^\mr{phys}$.}
\label{fig:fpifkmkme}
\end{figure}

In the 3-flavor case we have two independent quark masses, the average down and
up quark mass and the strange quark mass.
We take the liberty to rewrite everything in terms of $\Mpi$ and $m_s$, for
reasons that will become obvious soon.
Then the 1-loop quark mass formulae read
\begin{eqnarray}
\mk^2&=&\tree{\mk^2}+{1\ovr\fpi^2}
\bigg[
   \mpi^4\Big\{-2k_1+{1\ovr4N}(-\ell_\pi+{1\ovr3}\treel{\ell}_\eta)\Big\}
\nonumber\\
&&\hspace*{1.7cm}
   +m_sB_0(\mpi^2+m_sB_0)\Big\{8(k_1+2k_2)+{4\ovr9N}\treel{\ell}_\eta\Big\}
\bigg]
\label{mk}
\\
\me^2&=&\tree{\me^2}\,+{1\ovr\fpi^2}
\bigg[
\mpi^4\Big\{{16\ovr9}(-k_1+2k_3)+{1\ovr3N}(-2\ell_\pi+\treel{\ell}_K)\Big\}
\nonumber\\
&&\hspace*{1.7cm}+\mpi^2m_sB_0\Big\{{64\ovr9}(k_1+3k_2-2k_3)
+{4\ovr3N}(\treel{\ell}_K-{2\ovr9}\treel{\ell}_\eta)\Big\}
\nn&&\hspace*{1.7cm}
   +(m_sB_0)^2\Big\{{128\ovr9}(k_1+{3\ovr2}k_2+k_3)
   +{4\ovr3N}(\treel{\ell}_K-{8\ovr9}\treel{\ell}_\eta)\Big\}
\bigg]
\label{me}
\\
F_K&=&\fpi\,\,+{1\ovr\fpi}\,
\bigg[
4(\mk^2-\mpi^2)L_5^\mr{r}+{1\ovr N}\Big\{{5\ovr8}\mpi^2\ell_\pi
-{1\ovr4}\mk^2\ell_K-{3\ovr8}\me^2\ell_\eta\Big\}
\bigg]
\label{fk}
\end{eqnarray}
with
\bdm
k_1 = 2L_8^\mr{r}-L_5^\mr{r} \;,\qquad
k_2 = 2L_6^\mr{r}-L_4^\mr{r} \;,\qquad
k_3 = 3L_7^\mr{r}+L_8^\mr{r} 
\edm
and with the abbreviations
\beq
\tree{\mk^2}=m_s B_0 + {1\ovr2}\mpi^2 \;,\qquad
\tree{\me^2}={1\ovr3}(\mpi^2+4 m_s B_0)
\eeq
together with $N$ and $\ell_P$ as defined in (\ref{abbrev2}) and accordingly
$\treel{\ell}_P=\ln(\tree{M_P^2}/\mu^2)$.
Note that $\tree{\mk^2}$ and $\tree{\me^2}$ are of hybrid nature -- the
$\Mpi^2$ part refers to 1-loop ChPT, while the part linear in $m_s$ is a
tree-level contribution. This is unavoidable if we want to discuss the
dependence of physical quantities on the pion mass, instead of on quark
masses. In technical terms (\ref{mk}) is used to fix $m_sB_0$; we simply
require that $\mk$ takes the physical value for $\mpi\!=\!\mpi^\mr{phys}$,
the result is $B_0 m_s=0.223\mr{GeV}^2$.
The analogous requirement for $\me$ implies that we must slightly readjust
$L_7^\mr{r}$ to $-0.47\cdot10^{-3}$ [well compatible with the error in
Tab.\,\ref{tab:SU2SU3}], for all other low energy constants the central
values in Tab.\,\ref{tab:SU2SU3} are used. Even in the 3-flavor case we
choose to describe the quark mass dependence of $\mpi$ and $\fpi$ through
$SU(2)$ ChPT, as discussed in the previous subsection.
Note that for $m_s\!=\!m_s^\mr{phys}$ this choice exactly reproduces what one
would get in the $SU(3)$ framework, since the phenomenological $\bar\ell_i$
values know about the virtual strange quark loops. In actual lattice
simulations $m_s$ is typically close to the physical value, and we expect
this to remain a valid approximation. 
The resulting $\mpi$ dependence of $\fk,\mk,\me$ is shown in
Fig.\,\ref{fig:fpifkmkme}.
One notices that $\me\!\sim\!640\MeV$ for $\mpi\!\sim\!500\MeV$, thus the
expansion parameter $\xi_P$ remains small in the entire mass range
$\mpi\!\leq\!500\MeV$, even for $P\!=\!\et$.
In our numerical analysis we will use $\xi_P$ exactly as determined from
Fig.\,\ref{fig:fpifkmkme} and ignore the uncertainty of this computed expansion
parameter, since in a lattice computation one may iteratively determine
$M_P(L)$ and $\fpi(L)$ and thus $\xi_P$.
We do, however, consider the uncertainties in the expansion coefficients
$I_{M_P}^{(2/4/6)}$ and $I_{F_P}^{(2/4)}$ in (\ref{generic_M}) and
(\ref{generic_F}), respectively, with details specified below.
We have checked that even including the contribution of $\xi_P$ to the
total error would barely change the errors in our main result,
Figs.\,\ref{fig:Rmpi}-\ref{fig:Rmkaet}.
In summary, the quark mass dependence of a quantity $X\!=\!\fk,\mk,\me$
is considered a function of $\mpi$ alone through
\begin{equation}
X=X(\mpi,m_s B_0\!=\!0.223\GeV^2)
\end{equation}
and an appropriate choice of the $SU(3)$ low energy constants, as given in
Tab.\,\ref{tab:SU2SU3}.

\subsection{Results}

\begin{figure}
\centering
\includegraphics[width=15cm]{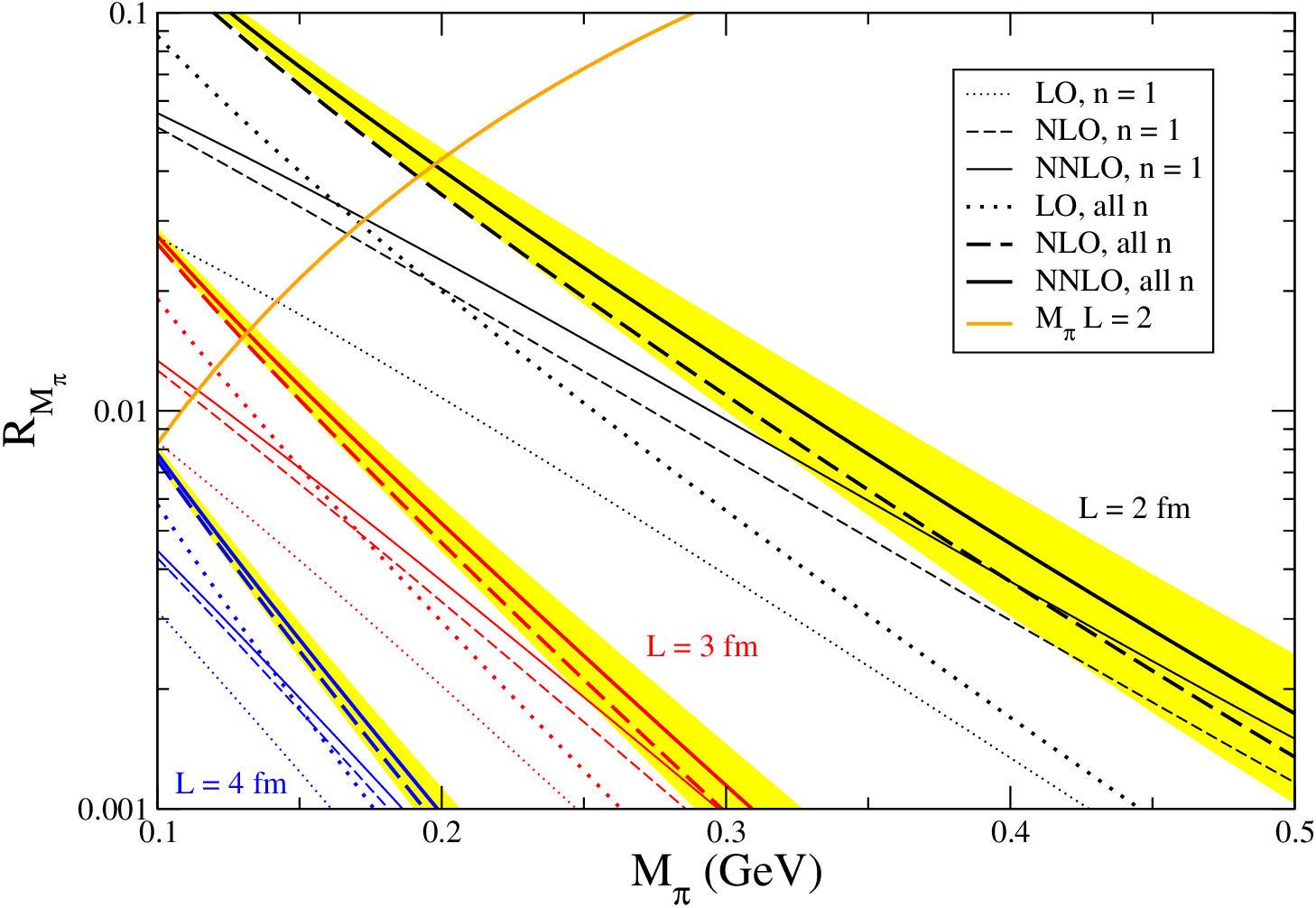}
\includegraphics[width=15cm]{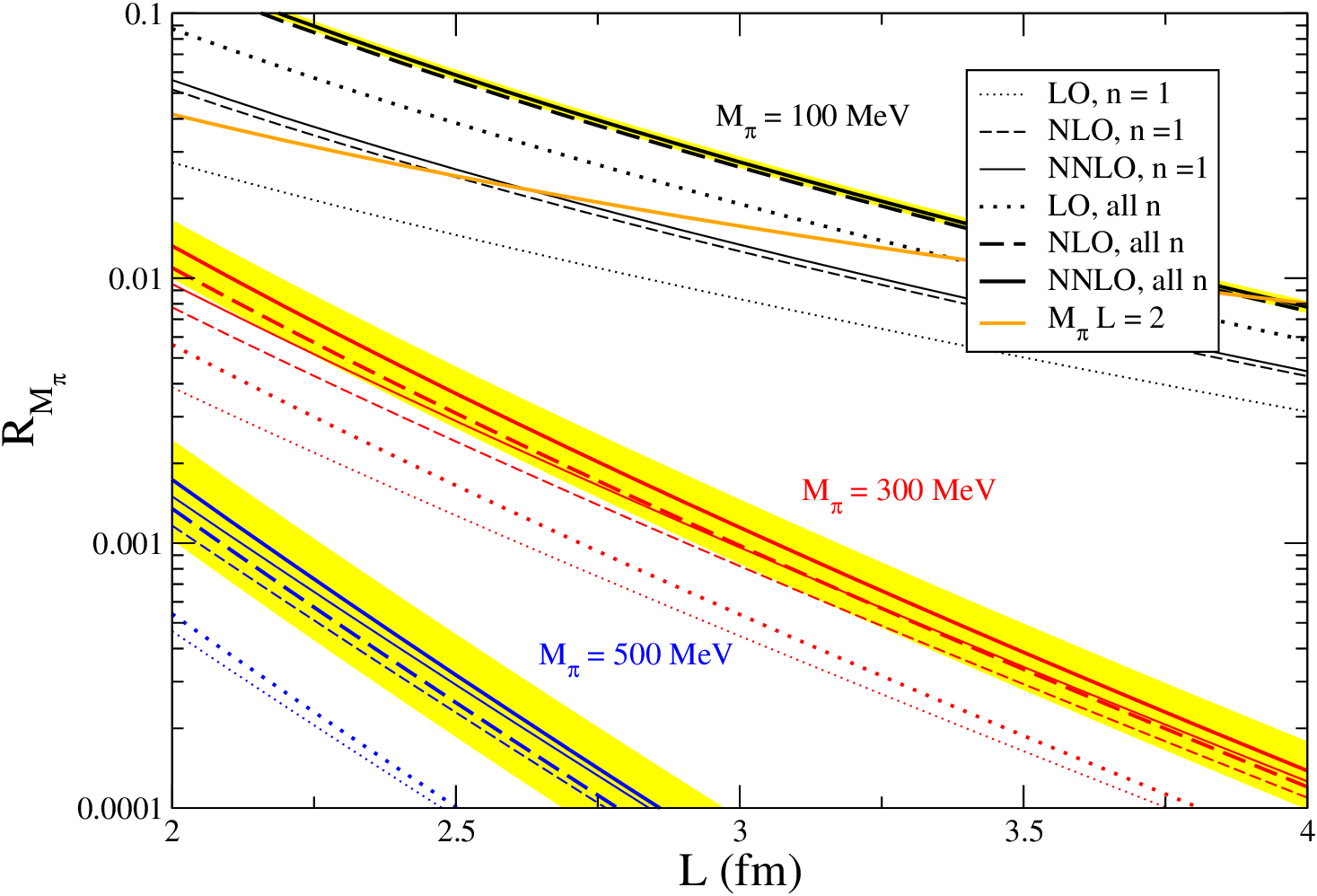}
\vspace{-2mm}
\caption{$R_{\mpi}$ vs.\ $\mpi$ for $L\!=\!2,3,4\fm$ (top) and vs.\ $L$
for $\mpi\!=\!100,300,500\MeV$ (bottom). The result of the original
(``$n\!=\!1$'') L\"uscher formula (\ref{luscher_mpi_ori}) with LO/NLO/NNLO
chiral input is to be compared to the resummed (``all $n$'') formula
(\ref{eq:Rmpin}) which amounts to an approximate 1/2/3-loop ChPT calculation
in finite volume. With NNLO input the low energy constants lead to
a non-negligible error band; with NLO input the error is smaller (not shown),
with LO input it is zero (see text). In the region above the $\mpi L\!=\!2$
line one is not safely in the $p$-regime and our results should not be
trusted.}
\label{fig:Rmpi}
\end{figure}

\begin{figure}
\centering
\includegraphics[width=8.3cm]{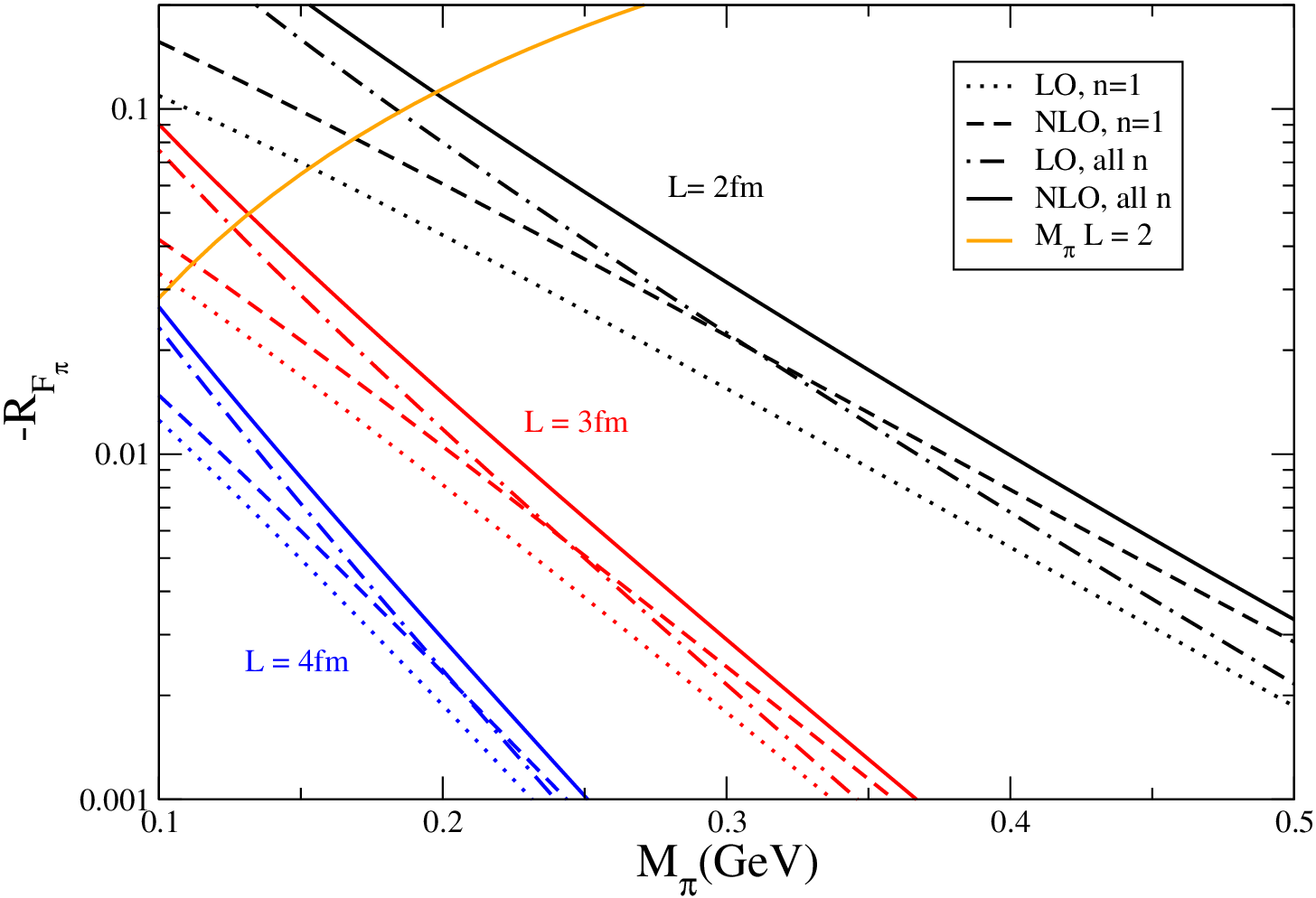}
\includegraphics[width=8.3cm]{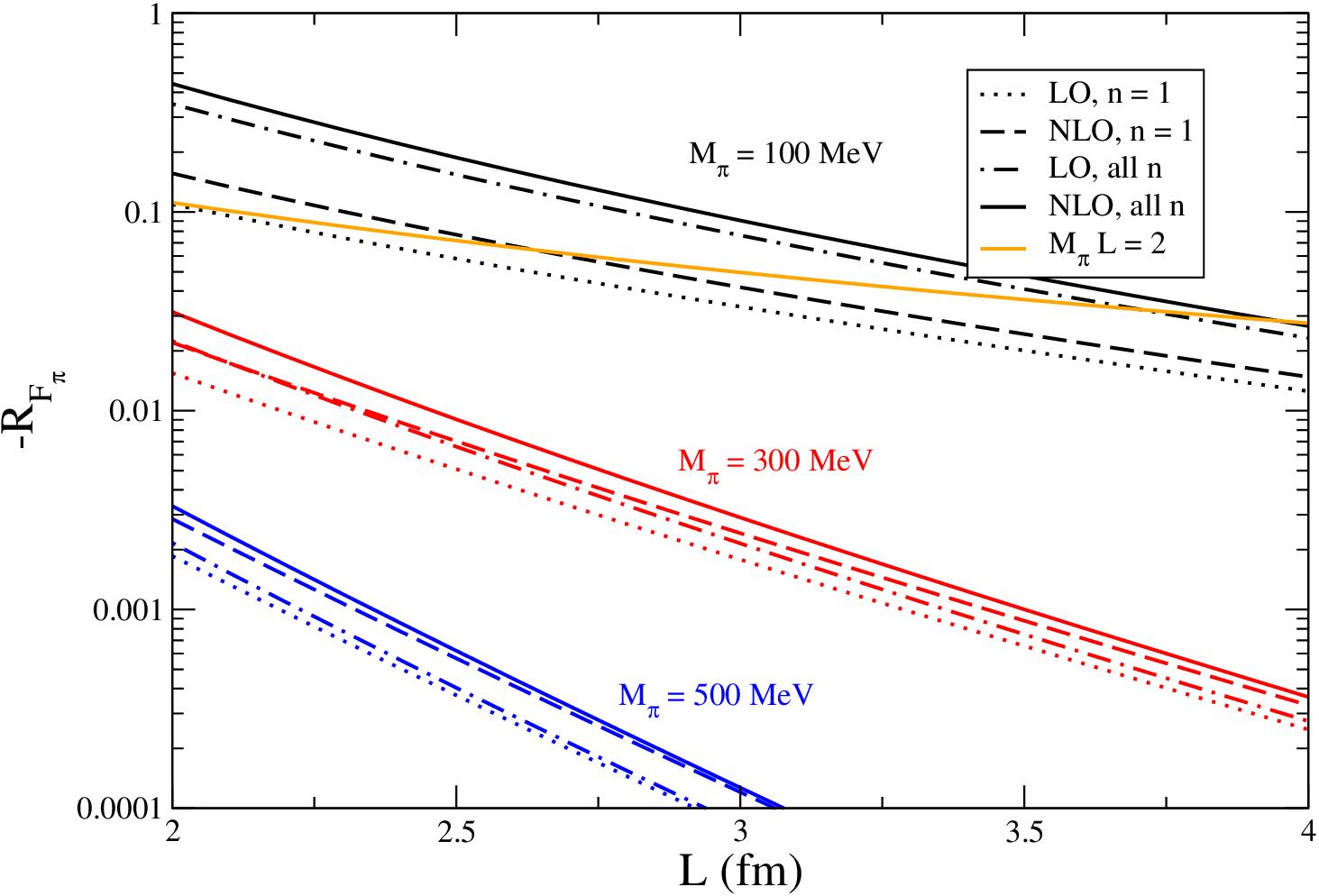}
\vspace{-2mm}
\caption{$-R_{\fpi}$ vs.\ $\mpi$ for $L\!=\!2,3,4\fm$ (left) and vs.\ $L$
for $\mpi\!=\!100,300,500\MeV$ (right).}
\label{fig:Rfpi}
\end{figure}

\begin{figure}
\centering
\includegraphics[width=8.3cm]{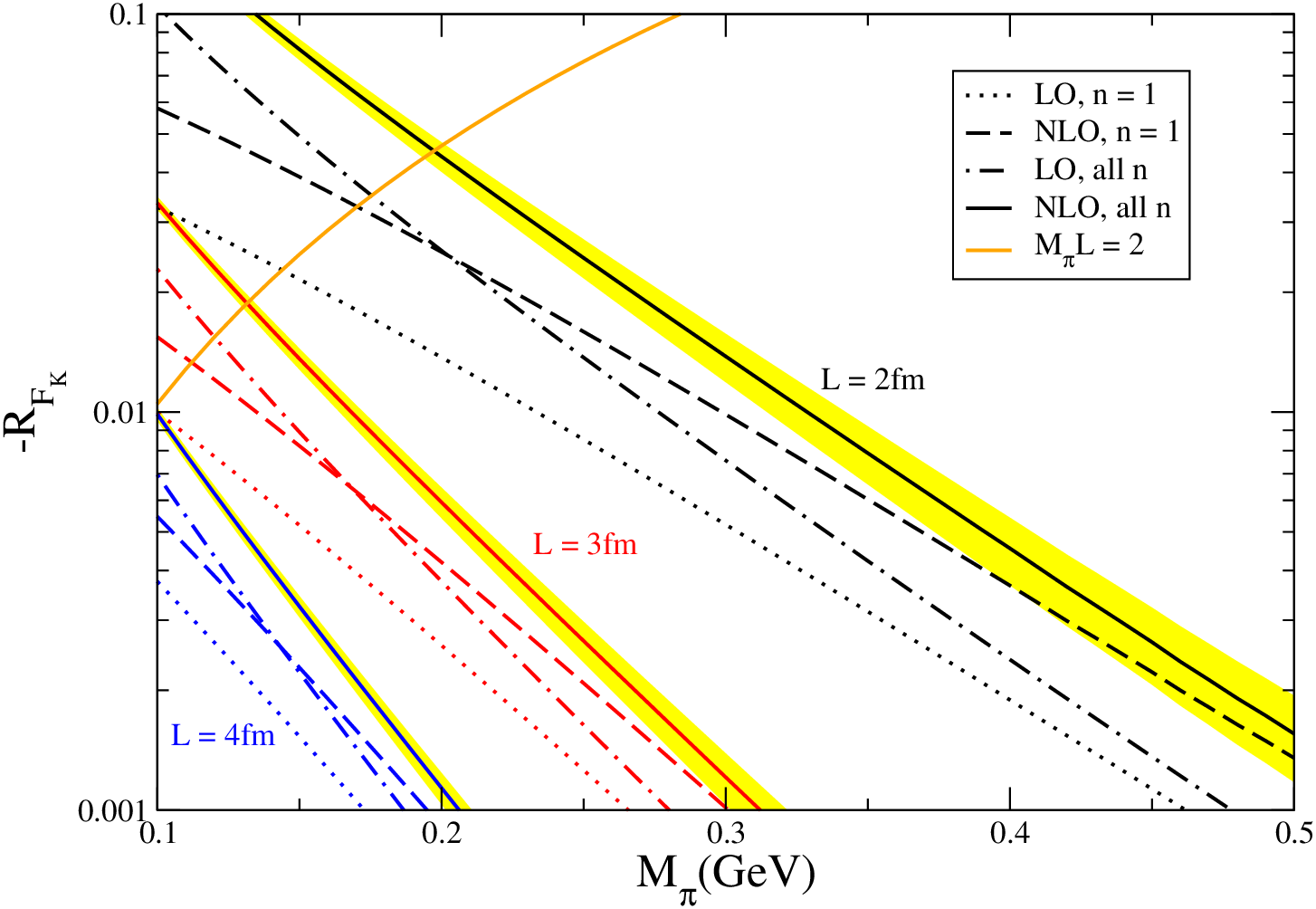}
\includegraphics[width=8.3cm]{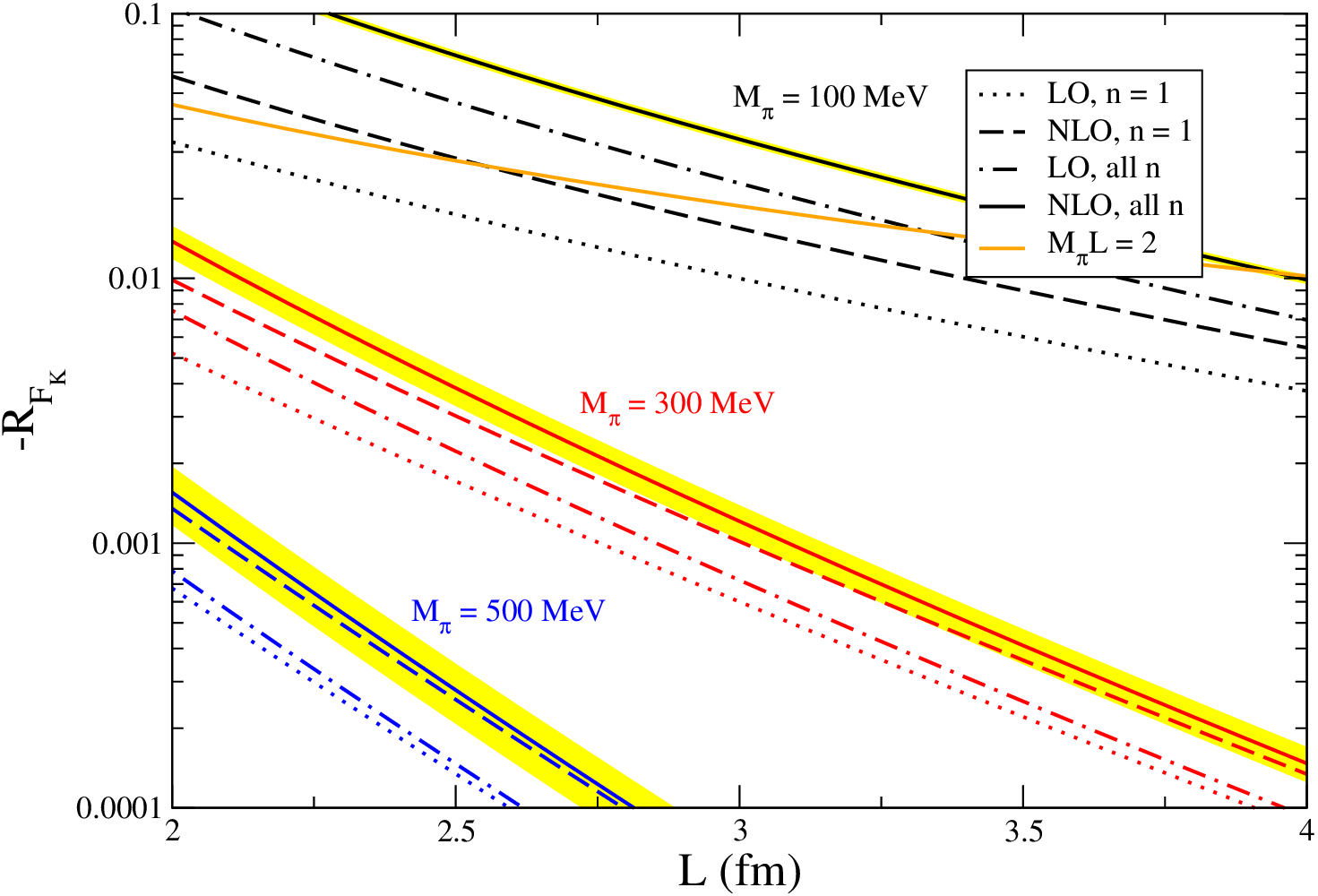}
\vspace{-2mm}
\caption{$-R_{\fk}$ vs.\ $\mpi$ for $L\!=\!2,3,4\fm$ (left) and vs.\ $L$
for $\mpi\!=\!100,300,500\MeV$ (right).}
\label{fig:Rfka}
\end{figure}

\begin{figure}
\centering
\includegraphics[width=8.3cm]{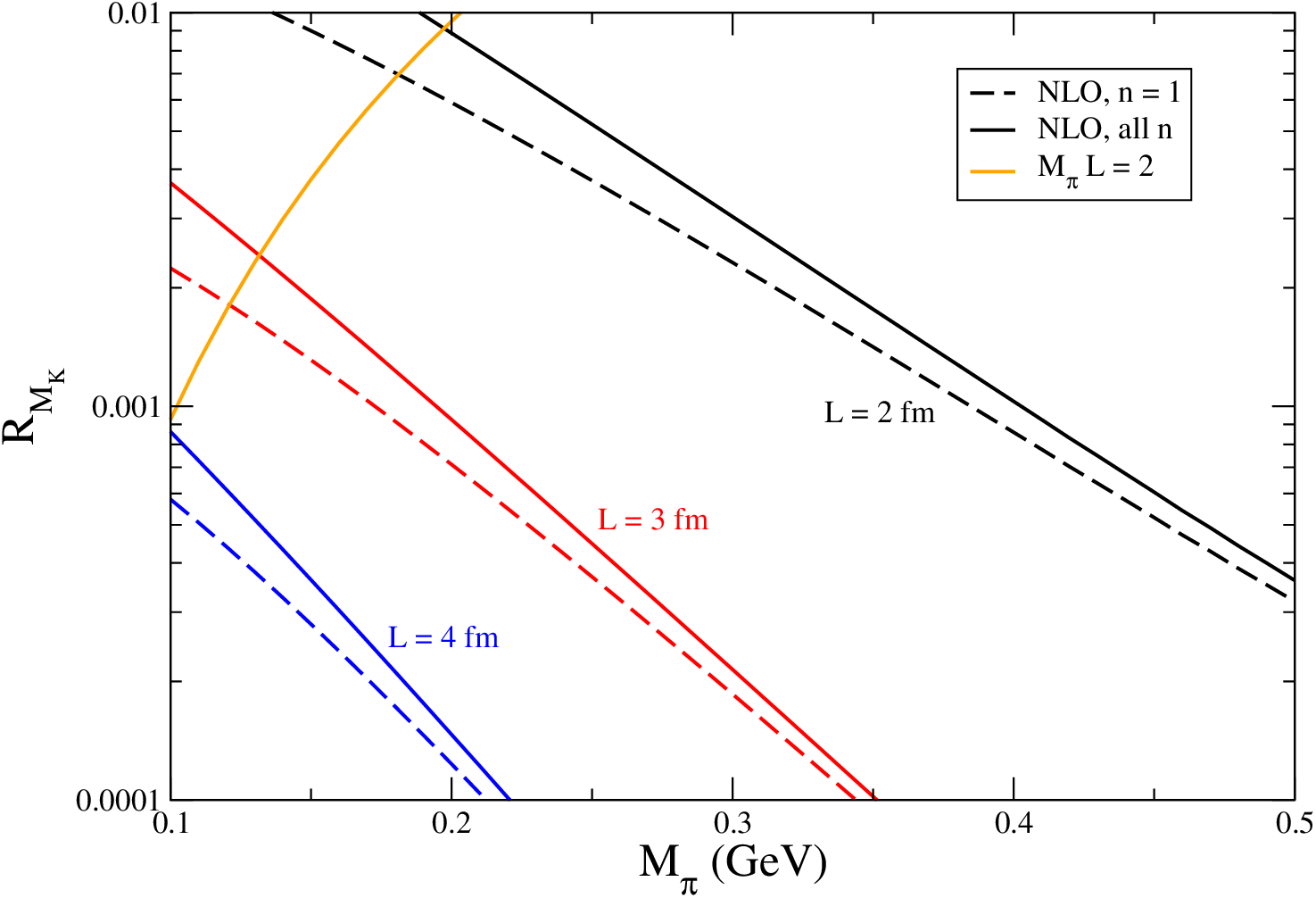}
\includegraphics[width=8.3cm]{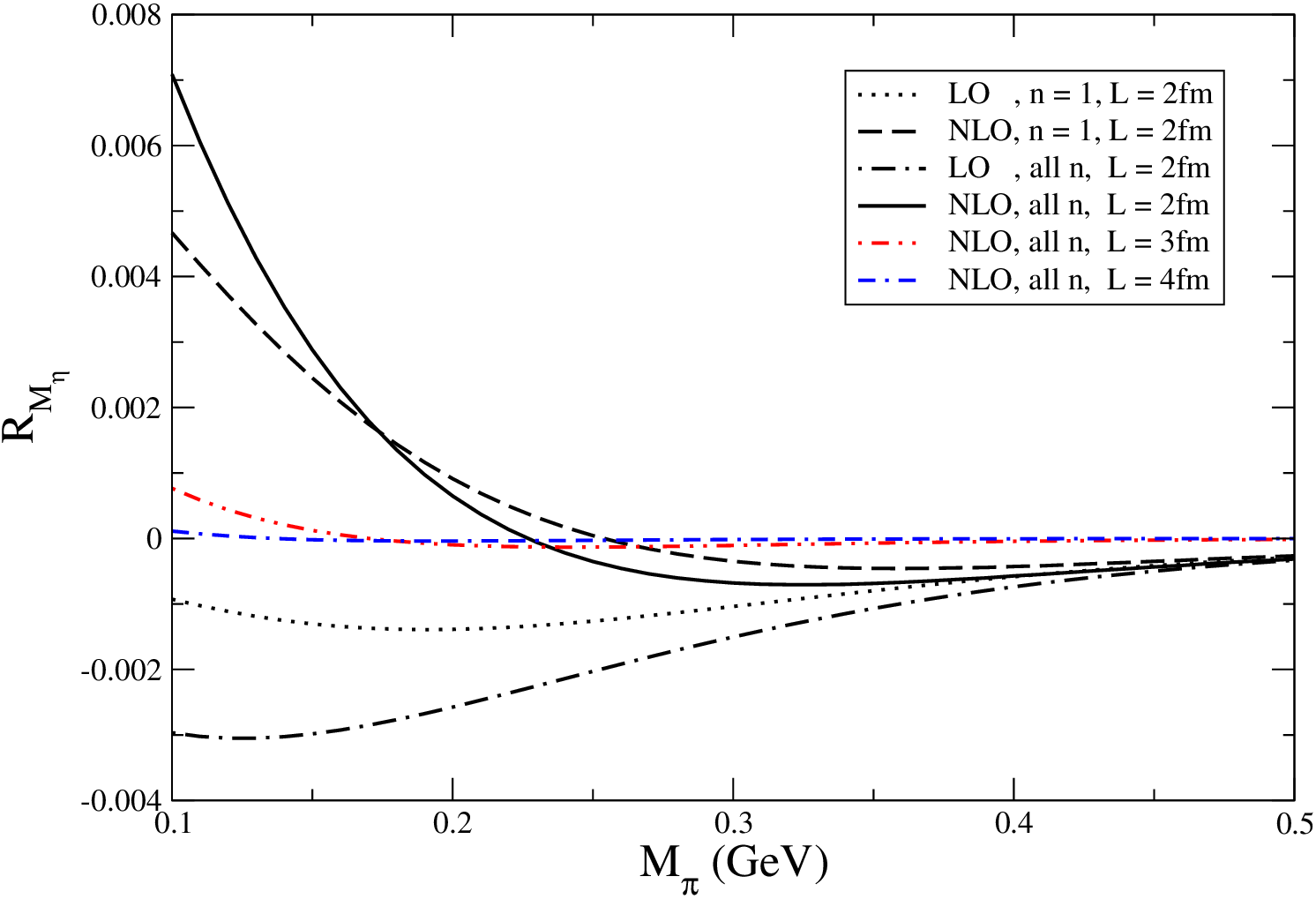}
\vspace{-2mm}
\caption{$R_{\mk}$ (left) and $R_{\me}$ (right) vs.\ $\mpi$ for
$L\!=\!2,3,4\fm$.}
\label{fig:Rmkaet}
\end{figure}

We plot our results for $R_{\mpi}$ in Fig.\,\ref{fig:Rmpi}, both for
$L\!=\!2,3,4\fm$ as a function of $\mpi$ (top) and for
$\mpi\!=\!100,300,500\MeV$ as a function of $L$ (bottom).
The result of the original (``$n\!=\!1$'') L\"uscher formula
(\ref{luscher_mpi_ori}) with LO/NLO/NNLO chiral input is shown with a thin
dotted/dashed/full line, respectively.
The resummed (``all $n$'') formula (\ref{eq:Rmpin}) with LO/NLO/NNLO chiral
input is given with a thick dotted/dashed/full line, respectively.
With NLO or NNLO input the pertinent low energy constants lead to a
non-negligible error band, except for $R_{\fpi}$ where the error with NLO
input is of the order of the thickness of the line, with LO input it is zero.
This is a consequence of our choice to disregard any uncertainty in the
expansion parameter (here $\xi_\pi$), as discussed in the previous subsection.
The area which corresponds --~in the resummed scenario with NNLO input~-- to
a situation with $\mpi L\!\leq\!2$ should be disregarded, since there is no
reason to hope that the resummed L\"uscher formula would still capture the
numerically dominating terms in a complete ``ChPT in finite volume'' formula
to the corresponding order in the $p$-counting.

Fig.\,\ref{fig:Rfpi} contains our data for $-R_{\fpi}$.
They are organized in the same manner as in the previous figure, though only
LO and NLO input is used, since the pertinent matrix element
is known only to 1-loop order, as discussed in Sect.\,4.

Finally, Figs.\,\ref{fig:Rfka}-\ref{fig:Rmkaet} contain the same information
for $\fk,\mk,\me$.
In all these cases the asymptotic formulae with and without resummation may
be compared, and the effect of going from LO to NLO input may be assessed.

\begin{table}[hp]
\footnotesize
\begin{tabular}{|c|cccccccc|}
\hline
$R_{M_\pi}$
&$1.6\fm$&$1.8\fm$&$2.0\fm$&$2.2\fm$&$2.4\fm$&$2.6\fm$&$2.8\fm$&$3.0\fm$\\
\hline
$140\MeV$&%
.2099(127)\hspace{-1mm}&\hspace{-1mm}%
0.1292(82)\hspace{-1mm}&\hspace{-1mm}%
0.0830(55)\hspace{-1mm}&\hspace{-1mm}%
0.0552(38)\hspace{-1mm}&\hspace{-1mm}%
0.0377(27)\hspace{-1mm}&\hspace{-1mm}%
0.0264(19)\hspace{-1mm}&\hspace{-1mm}%
0.0188(14)\hspace{-1mm}&\hspace{-1mm}%
0.0136(10)\\
$160\MeV$&%
.1687(130)\hspace{-1mm}&\hspace{-1mm}%
0.1023(83)\hspace{-1mm}&\hspace{-1mm}%
0.0647(55)\hspace{-1mm}&\hspace{-1mm}%
0.0424(37)\hspace{-1mm}&\hspace{-1mm}%
0.0285(26)\hspace{-1mm}&\hspace{-1mm}%
0.0196(18)\hspace{-1mm}&\hspace{-1mm}%
0.0138(13)\hspace{-1mm}&\hspace{-1mm}%
0.0098(10)\\
$180\MeV$&%
.1366(130)\hspace{-1mm}&\hspace{-1mm}%
0.0817(82)\hspace{-1mm}&\hspace{-1mm}%
0.0509(53)\hspace{-1mm}&\hspace{-1mm}%
0.0328(35)\hspace{-1mm}&\hspace{-1mm}%
0.0217(24)\hspace{-1mm}&\hspace{-1mm}%
0.0147(17)\hspace{-1mm}&\hspace{-1mm}%
0.0102(12)\hspace{-1mm}&\hspace{-1mm}%
0.0071(09)\\
$200\MeV$&%
.1113(127)\hspace{-1mm}&\hspace{-1mm}%
0.0656(79)\hspace{-1mm}&\hspace{-1mm}%
0.0403(50)\hspace{-1mm}&\hspace{-1mm}%
0.0256(33)\hspace{-1mm}&\hspace{-1mm}%
0.0167(22)\hspace{-1mm}&\hspace{-1mm}%
0.0111(15)\hspace{-1mm}&\hspace{-1mm}%
0.0076(11)\hspace{-1mm}&\hspace{-1mm}%
0.0052(08)\\
$220\MeV$&%
.0912(123)\hspace{-1mm}&\hspace{-1mm}%
0.0529(75)\hspace{-1mm}&\hspace{-1mm}%
0.0320(47)\hspace{-1mm}&\hspace{-1mm}%
0.0200(31)\hspace{-1mm}&\hspace{-1mm}%
0.0129(20)\hspace{-1mm}&\hspace{-1mm}%
0.0084(14)\hspace{-1mm}&\hspace{-1mm}%
0.0056(09)\hspace{-1mm}&\hspace{-1mm}%
0.0038(06)\\
$240\MeV$&%
.0749(117)\hspace{-1mm}&\hspace{-1mm}%
0.0429(70)\hspace{-1mm}&\hspace{-1mm}%
0.0256(43)\hspace{-1mm}&\hspace{-1mm}%
0.0157(28)\hspace{-1mm}&\hspace{-1mm}%
0.0099(18)\hspace{-1mm}&\hspace{-1mm}%
0.0064(12)\hspace{-1mm}&\hspace{-1mm}%
0.0042(08)\hspace{-1mm}&\hspace{-1mm}%
0.0028(05)\\
$260\MeV$&%
.0618(110)\hspace{-1mm}&\hspace{-1mm}%
0.0349(65)\hspace{-1mm}&\hspace{-1mm}%
0.0205(39)\hspace{-1mm}&\hspace{-1mm}%
0.0124(25)\hspace{-1mm}&\hspace{-1mm}%
0.0077(16)\hspace{-1mm}&\hspace{-1mm}%
0.0049(10)\hspace{-1mm}&\hspace{-1mm}%
0.0032(07)\hspace{-1mm}&\hspace{-1mm}%
0.0021(05)\\
$280\MeV$&%
.0511(102)\hspace{-1mm}&\hspace{-1mm}%
0.0284(59)\hspace{-1mm}&\hspace{-1mm}%
0.0164(35)\hspace{-1mm}&\hspace{-1mm}%
0.0098(22)\hspace{-1mm}&\hspace{-1mm}%
0.0060(14)\hspace{-1mm}&\hspace{-1mm}%
0.0038(09)\hspace{-1mm}&\hspace{-1mm}%
0.0024(06)\hspace{-1mm}&\hspace{-1mm}%
0.0015(04)\\
$300\MeV$&%
0.0423(93)\hspace{-1mm}&\hspace{-1mm}%
0.0232(53)\hspace{-1mm}&\hspace{-1mm}%
0.0132(31)\hspace{-1mm}&\hspace{-1mm}%
0.0078(19)\hspace{-1mm}&\hspace{-1mm}%
0.0047(12)\hspace{-1mm}&\hspace{-1mm}%
0.0029(07)\hspace{-1mm}&\hspace{-1mm}%
0.0018(05)\hspace{-1mm}&\hspace{-1mm}%
0.0011(03)\\
$320\MeV$&%
0.0352(85)\hspace{-1mm}&\hspace{-1mm}%
0.0190(48)\hspace{-1mm}&\hspace{-1mm}%
0.0107(28)\hspace{-1mm}&\hspace{-1mm}%
0.0062(16)\hspace{-1mm}&\hspace{-1mm}%
0.0037(10)\hspace{-1mm}&\hspace{-1mm}%
0.0022(06)\hspace{-1mm}&\hspace{-1mm}%
0.0014(04)\hspace{-1mm}&\hspace{-1mm}%
0.0009(02)\\
$340\MeV$&%
0.0293(77)\hspace{-1mm}&\hspace{-1mm}%
0.0156(42)\hspace{-1mm}&\hspace{-1mm}%
0.0086(24)\hspace{-1mm}&\hspace{-1mm}%
0.0049(14)\hspace{-1mm}&\hspace{-1mm}%
0.0029(08)\hspace{-1mm}&\hspace{-1mm}%
0.0017(05)\hspace{-1mm}&\hspace{-1mm}%
0.0010(03)\hspace{-1mm}&\hspace{-1mm}%
0.0006(02)\\
$360\MeV$&%
0.0245(69)\hspace{-1mm}&\hspace{-1mm}%
0.0129(37)\hspace{-1mm}&\hspace{-1mm}%
0.0070(21)\hspace{-1mm}&\hspace{-1mm}%
0.0039(12)\hspace{-1mm}&\hspace{-1mm}%
0.0023(07)\hspace{-1mm}&\hspace{-1mm}%
0.0013(04)\hspace{-1mm}&\hspace{-1mm}%
0.0008(03)\hspace{-1mm}&\hspace{-1mm}%
0.0005(02)\\
$380\MeV$&%
0.0205(62)\hspace{-1mm}&\hspace{-1mm}%
0.0106(33)\hspace{-1mm}&\hspace{-1mm}%
0.0057(18)\hspace{-1mm}&\hspace{-1mm}%
0.0031(10)\hspace{-1mm}&\hspace{-1mm}%
0.0018(06)\hspace{-1mm}&\hspace{-1mm}%
0.0010(03)\hspace{-1mm}&\hspace{-1mm}%
0.0006(02)\hspace{-1mm}&\hspace{-1mm}%
0.0004(01)\\
$400\MeV$&%
0.0172(55)\hspace{-1mm}&\hspace{-1mm}%
0.0088(29)\hspace{-1mm}&\hspace{-1mm}%
0.0046(15)\hspace{-1mm}&\hspace{-1mm}%
0.0025(09)\hspace{-1mm}&\hspace{-1mm}%
0.0014(05)\hspace{-1mm}&\hspace{-1mm}%
0.0008(03)\hspace{-1mm}&\hspace{-1mm}%
0.0005(02)\hspace{-1mm}&\hspace{-1mm}%
0.0003(01)\\
$420\MeV$&%
0.0145(49)\hspace{-1mm}&\hspace{-1mm}%
0.0073(25)\hspace{-1mm}&\hspace{-1mm}%
0.0038(13)\hspace{-1mm}&\hspace{-1mm}%
0.0020(07)\hspace{-1mm}&\hspace{-1mm}%
0.0011(04)\hspace{-1mm}&\hspace{-1mm}%
0.0006(02)\hspace{-1mm}&\hspace{-1mm}%
0.0003(01)\hspace{-1mm}&\hspace{-1mm}%
0.0002(01)\\
$440\MeV$&%
0.0123(43)\hspace{-1mm}&\hspace{-1mm}%
0.0061(22)\hspace{-1mm}&\hspace{-1mm}%
0.0031(11)\hspace{-1mm}&\hspace{-1mm}%
0.0016(06)\hspace{-1mm}&\hspace{-1mm}%
0.0009(03)\hspace{-1mm}&\hspace{-1mm}%
0.0005(02)\hspace{-1mm}&\hspace{-1mm}%
0.0003(01)\hspace{-1mm}&\hspace{-1mm}%
0.0001(01)\\
$460\MeV$&%
0.0104(38)\hspace{-1mm}&\hspace{-1mm}%
0.0051(19)\hspace{-1mm}&\hspace{-1mm}%
0.0025(10)\hspace{-1mm}&\hspace{-1mm}%
0.0013(05)\hspace{-1mm}&\hspace{-1mm}%
0.0007(03)\hspace{-1mm}&\hspace{-1mm}%
0.0004(01)\hspace{-1mm}&\hspace{-1mm}%
0.0002(01)\hspace{-1mm}&\hspace{-1mm}%
0.0001(00)\\
$480\MeV$&%
0.0088(33)\hspace{-1mm}&\hspace{-1mm}%
0.0042(16)\hspace{-1mm}&\hspace{-1mm}%
0.0021(08)\hspace{-1mm}&\hspace{-1mm}%
0.0011(04)\hspace{-1mm}&\hspace{-1mm}%
0.0006(02)\hspace{-1mm}&\hspace{-1mm}%
0.0003(01)\hspace{-1mm}&\hspace{-1mm}%
0.0002(01)\hspace{-1mm}&\hspace{-1mm}%
0.0001(00)\\
$500\MeV$&%
0.0076(29)\hspace{-1mm}&\hspace{-1mm}%
0.0036(14)\hspace{-1mm}&\hspace{-1mm}%
0.0017(07)\hspace{-1mm}&\hspace{-1mm}%
0.0009(03)\hspace{-1mm}&\hspace{-1mm}%
0.0004(02)\hspace{-1mm}&\hspace{-1mm}%
0.0002(01)\hspace{-1mm}&\hspace{-1mm}%
0.0001(00)\hspace{-1mm}&\hspace{-1mm}%
0.0001(00)\\
\hline
\end{tabular}
\normalsize
\vspace{-2mm}
\caption{$R_{\Mpi}$ via the resummed L\"uscher formula (\ref{RMP}) with NNLO
chiral input for $\cF_{\pi}(\til\nu)$, representing an approximate 3-loop
result. The error includes the uncertainty of the $\bar\ell_i$ and the
$O(p^6)$ low energy constants, but no systematics. Entries with $\Mpi
L\!<\!2$ are unlikely to really capture the physical finite size effect, and
the first two columns are somewhat on the short side with respect to the
condition (\ref{xpt_cond1}).} 
\label{tab:Rmpi}
\vspace{+6mm}
\footnotesize
\begin{tabular}{|c|cccccccc|}
\hline
$-R_{F_\pi}$
&$1.6\fm$&$1.8\fm$&$2.0\fm$&$2.2\fm$&$2.4\fm$&$2.6\fm$&$2.8\fm$&$3.0\fm$\\
\hline
$140\MeV$&%
0.5844(43)\hspace{-1mm}&\hspace{-1mm}%
0.3683(24)\hspace{-1mm}&\hspace{-1mm}%
0.2414(14)\hspace{-1mm}&\hspace{-1mm}%
0.1633(09)\hspace{-1mm}&\hspace{-1mm}%
0.1134(06)\hspace{-1mm}&\hspace{-1mm}%
0.0804(04)\hspace{-1mm}&\hspace{-1mm}%
0.0581(03)\hspace{-1mm}&\hspace{-1mm}%
0.0426(02)\\
$160\MeV$&%
0.4551(36)\hspace{-1mm}&\hspace{-1mm}%
0.2828(20)\hspace{-1mm}&\hspace{-1mm}%
0.1827(12)\hspace{-1mm}&\hspace{-1mm}%
0.1218(07)\hspace{-1mm}&\hspace{-1mm}%
0.0833(05)\hspace{-1mm}&\hspace{-1mm}%
0.0581(03)\hspace{-1mm}&\hspace{-1mm}%
0.0413(02)\hspace{-1mm}&\hspace{-1mm}%
0.0298(02)\\
$180\MeV$&%
0.3580(30)\hspace{-1mm}&\hspace{-1mm}%
0.2194(17)\hspace{-1mm}&\hspace{-1mm}%
0.1397(10)\hspace{-1mm}&\hspace{-1mm}%
0.0917(06)\hspace{-1mm}&\hspace{-1mm}%
0.0618(04)\hspace{-1mm}&\hspace{-1mm}%
0.0424(03)\hspace{-1mm}&\hspace{-1mm}%
0.0297(02)\hspace{-1mm}&\hspace{-1mm}%
0.0211(01)\\
$200\MeV$&%
0.2840(25)\hspace{-1mm}&\hspace{-1mm}%
0.1715(14)\hspace{-1mm}&\hspace{-1mm}%
0.1076(08)\hspace{-1mm}&\hspace{-1mm}%
0.0696(05)\hspace{-1mm}&\hspace{-1mm}%
0.0461(03)\hspace{-1mm}&\hspace{-1mm}%
0.0312(02)\hspace{-1mm}&\hspace{-1mm}%
0.0215(02)\hspace{-1mm}&\hspace{-1mm}%
0.0150(01)\\
$220\MeV$&%
0.2267(22)\hspace{-1mm}&\hspace{-1mm}%
0.1350(12)\hspace{-1mm}&\hspace{-1mm}%
0.0834(07)\hspace{-1mm}&\hspace{-1mm}%
0.0531(05)\hspace{-1mm}&\hspace{-1mm}%
0.0347(03)\hspace{-1mm}&\hspace{-1mm}%
0.0231(02)\hspace{-1mm}&\hspace{-1mm}%
0.0156(01)\hspace{-1mm}&\hspace{-1mm}%
0.0107(01)\\
$240\MeV$&%
0.1820(19)\hspace{-1mm}&\hspace{-1mm}%
0.1068(11)\hspace{-1mm}&\hspace{-1mm}%
0.0650(06)\hspace{-1mm}&\hspace{-1mm}%
0.0408(04)\hspace{-1mm}&\hspace{-1mm}%
0.0262(03)\hspace{-1mm}&\hspace{-1mm}%
0.0171(02)\hspace{-1mm}&\hspace{-1mm}%
0.0114(01)\hspace{-1mm}&\hspace{-1mm}%
0.0077(01)\\
$260\MeV$&%
0.1467(16)\hspace{-1mm}&\hspace{-1mm}%
0.0848(09)\hspace{-1mm}&\hspace{-1mm}%
0.0509(05)\hspace{-1mm}&\hspace{-1mm}%
0.0314(03)\hspace{-1mm}&\hspace{-1mm}%
0.0198(02)\hspace{-1mm}&\hspace{-1mm}%
0.0128(01)\hspace{-1mm}&\hspace{-1mm}%
0.0084(01)\hspace{-1mm}&\hspace{-1mm}%
0.0056(01)\\
$280\MeV$&%
0.1188(14)\hspace{-1mm}&\hspace{-1mm}%
0.0677(08)\hspace{-1mm}&\hspace{-1mm}%
0.0399(05)\hspace{-1mm}&\hspace{-1mm}%
0.0243(03)\hspace{-1mm}&\hspace{-1mm}%
0.0151(02)\hspace{-1mm}&\hspace{-1mm}%
0.0095(01)\hspace{-1mm}&\hspace{-1mm}%
0.0061(01)\hspace{-1mm}&\hspace{-1mm}%
0.0040(01)\\
$300\MeV$&%
0.0965(13)\hspace{-1mm}&\hspace{-1mm}%
0.0541(07)\hspace{-1mm}&\hspace{-1mm}%
0.0315(04)\hspace{-1mm}&\hspace{-1mm}%
0.0188(03)\hspace{-1mm}&\hspace{-1mm}%
0.0115(02)\hspace{-1mm}&\hspace{-1mm}%
0.0072(01)\hspace{-1mm}&\hspace{-1mm}%
0.0045(01)\hspace{-1mm}&\hspace{-1mm}%
0.0029(00)\\
$320\MeV$&%
0.0786(11)\hspace{-1mm}&\hspace{-1mm}%
0.0434(06)\hspace{-1mm}&\hspace{-1mm}%
0.0249(04)\hspace{-1mm}&\hspace{-1mm}%
0.0146(02)\hspace{-1mm}&\hspace{-1mm}%
0.0088(01)\hspace{-1mm}&\hspace{-1mm}%
0.0054(01)\hspace{-1mm}&\hspace{-1mm}%
0.0033(01)\hspace{-1mm}&\hspace{-1mm}%
0.0021(00)\\
$340\MeV$&%
0.0642(10)\hspace{-1mm}&\hspace{-1mm}%
0.0350(05)\hspace{-1mm}&\hspace{-1mm}%
0.0197(03)\hspace{-1mm}&\hspace{-1mm}%
0.0114(02)\hspace{-1mm}&\hspace{-1mm}%
0.0067(01)\hspace{-1mm}&\hspace{-1mm}%
0.0040(01)\hspace{-1mm}&\hspace{-1mm}%
0.0025(00)\hspace{-1mm}&\hspace{-1mm}%
0.0015(00)\\
$360\MeV$&%
0.0527(09)\hspace{-1mm}&\hspace{-1mm}%
0.0282(05)\hspace{-1mm}&\hspace{-1mm}%
0.0156(03)\hspace{-1mm}&\hspace{-1mm}%
0.0089(02)\hspace{-1mm}&\hspace{-1mm}%
0.0052(01)\hspace{-1mm}&\hspace{-1mm}%
0.0031(01)\hspace{-1mm}&\hspace{-1mm}%
0.0018(00)\hspace{-1mm}&\hspace{-1mm}%
0.0011(00)\\
$380\MeV$&%
0.0433(08)\hspace{-1mm}&\hspace{-1mm}%
0.0228(04)\hspace{-1mm}&\hspace{-1mm}%
0.0124(02)\hspace{-1mm}&\hspace{-1mm}%
0.0070(01)\hspace{-1mm}&\hspace{-1mm}%
0.0040(01)\hspace{-1mm}&\hspace{-1mm}%
0.0023(00)\hspace{-1mm}&\hspace{-1mm}%
0.0014(00)\hspace{-1mm}&\hspace{-1mm}%
0.0008(00)\\
$400\MeV$&%
0.0356(07)\hspace{-1mm}&\hspace{-1mm}%
0.0185(04)\hspace{-1mm}&\hspace{-1mm}%
0.0099(02)\hspace{-1mm}&\hspace{-1mm}%
0.0055(01)\hspace{-1mm}&\hspace{-1mm}%
0.0031(01)\hspace{-1mm}&\hspace{-1mm}%
0.0017(00)\hspace{-1mm}&\hspace{-1mm}%
0.0010(00)\hspace{-1mm}&\hspace{-1mm}%
0.0006(00)\\
$420\MeV$&%
0.0294(06)\hspace{-1mm}&\hspace{-1mm}%
0.0150(03)\hspace{-1mm}&\hspace{-1mm}%
0.0079(02)\hspace{-1mm}&\hspace{-1mm}%
0.0043(01)\hspace{-1mm}&\hspace{-1mm}%
0.0024(01)\hspace{-1mm}&\hspace{-1mm}%
0.0013(00)\hspace{-1mm}&\hspace{-1mm}%
0.0008(00)\hspace{-1mm}&\hspace{-1mm}%
0.0004(00)\\
$440\MeV$&%
0.0244(05)\hspace{-1mm}&\hspace{-1mm}%
0.0122(03)\hspace{-1mm}&\hspace{-1mm}%
0.0063(01)\hspace{-1mm}&\hspace{-1mm}%
0.0034(01)\hspace{-1mm}&\hspace{-1mm}%
0.0018(00)\hspace{-1mm}&\hspace{-1mm}%
0.0010(00)\hspace{-1mm}&\hspace{-1mm}%
0.0006(00)\hspace{-1mm}&\hspace{-1mm}%
0.0003(00)\\
$460\MeV$&%
0.0202(05)\hspace{-1mm}&\hspace{-1mm}%
0.0100(02)\hspace{-1mm}&\hspace{-1mm}%
0.0051(01)\hspace{-1mm}&\hspace{-1mm}%
0.0027(01)\hspace{-1mm}&\hspace{-1mm}%
0.0014(00)\hspace{-1mm}&\hspace{-1mm}%
0.0008(00)\hspace{-1mm}&\hspace{-1mm}%
0.0004(00)\hspace{-1mm}&\hspace{-1mm}%
0.0002(00)\\
$480\MeV$&%
0.0169(04)\hspace{-1mm}&\hspace{-1mm}%
0.0082(02)\hspace{-1mm}&\hspace{-1mm}%
0.0041(01)\hspace{-1mm}&\hspace{-1mm}%
0.0021(01)\hspace{-1mm}&\hspace{-1mm}%
0.0011(00)\hspace{-1mm}&\hspace{-1mm}%
0.0006(00)\hspace{-1mm}&\hspace{-1mm}%
0.0003(00)\hspace{-1mm}&\hspace{-1mm}%
0.0002(00)\\
$500\MeV$&%
0.0141(04)\hspace{-1mm}&\hspace{-1mm}%
0.0067(02)\hspace{-1mm}&\hspace{-1mm}%
0.0033(01)\hspace{-1mm}&\hspace{-1mm}%
0.0017(00)\hspace{-1mm}&\hspace{-1mm}%
0.0009(00)\hspace{-1mm}&\hspace{-1mm}%
0.0004(00)\hspace{-1mm}&\hspace{-1mm}%
0.0002(00)\hspace{-1mm}&\hspace{-1mm}%
0.0001(00)\\
\hline
\end{tabular}
\normalsize
\vspace{-2mm}
\caption{$-R_{\Fpi}$ via the resummed L\"uscher formula (\ref{RFP}) with NLO
chiral input for $\cN_{\pi}(\til\nu)$, representing an approximate 2-loop
result. The error includes the uncertainty of the $\bar\ell_i$, but no
systematics. Entries with $\Mpi L\!<\!2$ are unlikely to really capture the
physical finite size effect, and the first two columns are somewhat on the
short side with respect to the condition (\ref{xpt_cond1}).}
\label{tab:Rfpi}
\end{table}

The main message to be extracted from Figs.\,\ref{fig:Rmpi}-\ref{fig:Rmkaet}
is that the relative finite volume shift vanishes indeed in proportion to
$e^{-\Mpi L}$; in the logarithmic representation one has an almost linear
fall-off pattern, and higher orders mainly affect the prefactor.
The second point concerns the relative importance of higher orders
in the chiral counting versus higher exponentials.
For small pion masses (say $100\MeV$) resumming (i.e.\ higher exponentials)
prove vital, while for large pion masses (say $500\MeV$) higher loop orders
prove more relevant (though the effect is small in that regime).
As has been discussed in Ref.\,\cite{Colangelo:2003hf} a large shift in
$R_{M_P}$, when moving from LO to NLO input, does not necessarily signal a
bad chiral convergence behavior, since the cut in the underlying $\cF_P$
amplitude starts only at the NLO level.
Unfortunately, the effect due to an upgrade to NNLO input can only be checked
in the case of $R_{\mpi}$, since only there the pertinent amplitude is known.
Nonetheless, we believe that the regime in the $(\mpi,L)$ plane that leads to
a nice convergence behavior in $R_{\mpi}$ is indicative of the regime
where the resummed formulae for $R_{\fpi}, R_{\mk}, R_{\fk}, R_{\me}$ with
NLO input yield a trustworthy result.

For $R_{\mpi}$ and $R_{\fpi}$ the numerical results to the highest loop
order available (equivalent to an approximate 3-loop and 2-loop calculation
in ChPT) have been collected in Tabs.\,\ref{tab:Rmpi} and \ref{tab:Rfpi},
respectively.  From the general discussion it is clear that the logarithms
of the numbers in these tables may be interpolated with a low-order
polynomial.

 
\section{Two types of applications}


We finish with a discussion of two prototype applications of our formulae.
The first one is a ``forward-type'' application, in which our formulae are
used to control a systematic error in a lattice calculation.  The second
one concerns a ``backward-type'' application, where one tries to determine
QCD low energy constants from explicitly measuring finite-volume effects. 


\subsection{Finite volume effects and Marciano's determination of $V_{us}$}

An example of how an analytic finite volume calculation may help to
control a systematic error is the following.
Marciano pointed out that, modulo radiative corrections, the ratio
$\frac{V_{us}}{V_{ud}} \frac{F_K}{F_\pi}$ is fixed by the ratio of
branching ratios for $K_{\ell 2}$ and $\pi_{\ell 2}$ decays. Taking into
account radiative corrections, he obtained the following relation
\cite{Marciano:2004uf}
\begin{equation}
{|V_{us}|^2\ovr|V_{ud}|^2} {F_K^2\ovr F_\pi^2} = 0.07602(23)(27)
\end{equation}
where the errors represent the experimental and radiative correction
uncertainties. He then suggested to combine the value for $V_{ud}$
obtained from superallowed nuclear beta decays with a value for $\fk/\fpi$
from lattice simulations. We stress that the necessary accuracy to make an
impact on the determination of $V_{us}$ is at the level of 1\% or better,
and indeed, both the determination of $V_{ud}$ as well as the ratio of
branching ratios are known to well below 1\%. This means that any
improvement in the lattice calculation of $\fk/\fpi$ will be immediately
reflected in the value of $V_{us}$. In particular, being able to control
systematic effects to well below 1\% is of crucial importance.

\begin{figure}[t]
\centering
\includegraphics[width=15cm]{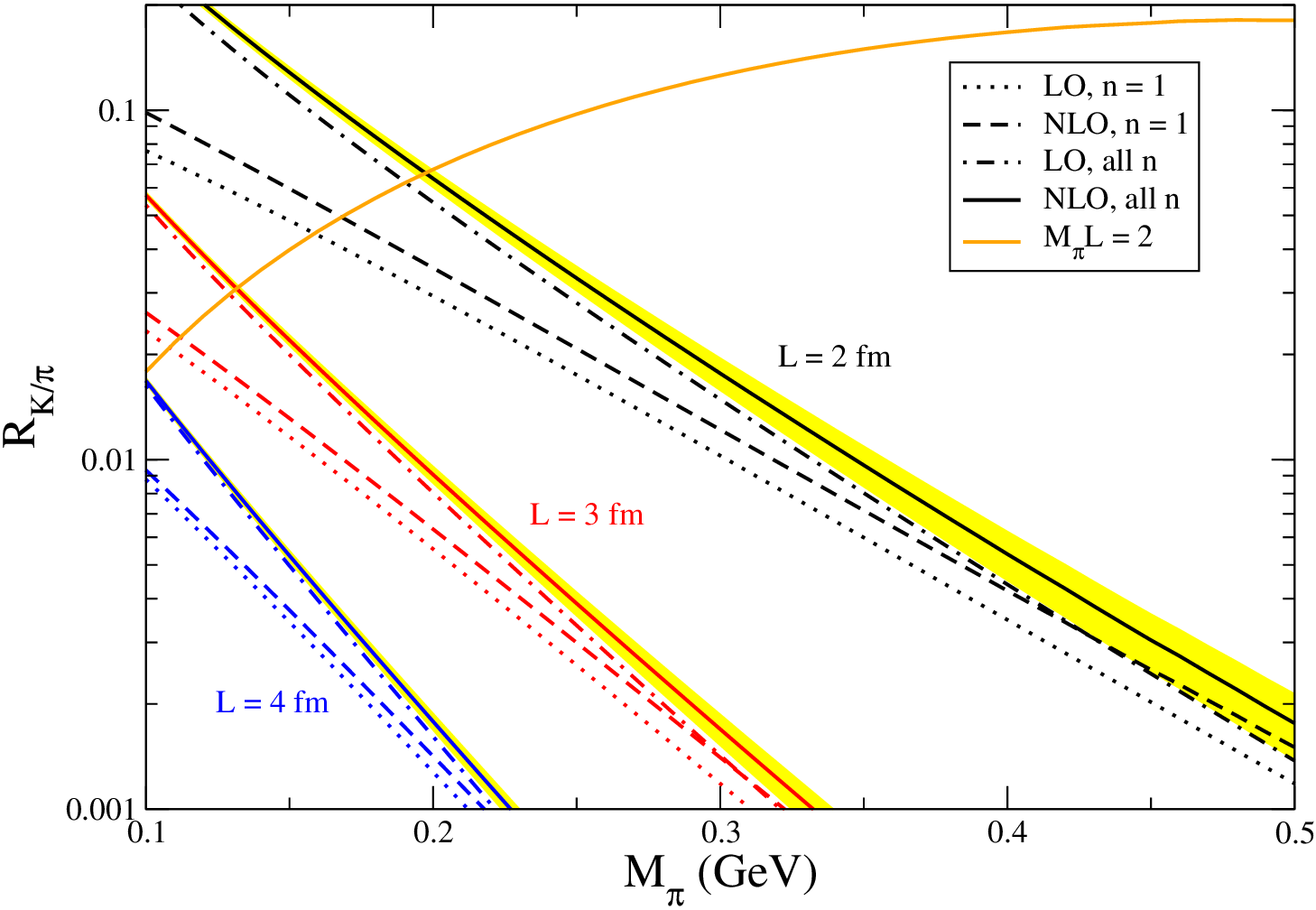}
\vspace{-2mm}
\caption{The relative finite volume effect $(\fk(L)\fpi)/(\fpi(L)\fk)-1$ vs.\
$\mpi$ for $L\!=\!2,3,4\fm$. Above the $\mpi L\!=\!2$ line one is not safely in
the $p$-regime and our results should not be trusted.}
\label{fig:Rfpifka}
\end{figure}

With our results for $R_{\fpi}$ and $R_{\fk}$ it is straightforward to
calculate the finite-volume shift of the latter ratio
\beq
{F_{K}(L)\ovr F_{\pi}(L)} = {F_K \ovr F_\pi}\;
\Big\{ 1+ R_{\fk} - R_{\fpi} + \mathcal{O}(R_F^2)\Big\}
\label{R_fk_over_fpi}
\eeq
and thus to compare the magnitude of this effect to the typical size of the
statistical error. A plot of the finite volume effect for the ratio of
decay constants as a function of the pion mass and for a few volume sizes
is provided in Fig.~\ref{fig:Rfpifka}.

In his analysis Marciano uses the MILC Collaboration result
$\fk/\fpi=1.201(8)(15)$ \cite{Aubin:2003ne,Aubin:2004wf}.
Among the various data sets they have, those with the smallest $M_\pi L$
and thus most likely to be affected by sizeable finite volume corrections,
have $\Mpi(L)\!\simeq\!311\MeV$, $L\!\simeq\!2.4\fm$ and
$\Mpi(L)\!\simeq\!262\MeV$, $L\!\simeq\!2.89\fm$.  Using
(\ref{R_fk_over_fpi}) we find that with the parameters of the first set
$\fk(L)/\fpi(L)$ {\em in the continuum} deviates from the infinite volume
result by $0.0099-0.0038=0.61$\%, while the corresponding estimate for the
second set reads $0.0067-0.0025=0.42$\%.  A systematic effect of half a
percent needs to be taken into account in a high precision study, and from
Fig.\,\ref{fig:Rfpifka} one sees that other ($\mpi,L$) pairs to reach that
level would be ($410\MeV,2\fm$), ($230\MeV,3\fm$) and ($150\MeV,4\fm$).

We stress that the numerical example just discussed is for illustrative
purposes only, because we do not know whether our formulae can be applied
to the MILC Collaboration data. 
Lattice QCD with staggered fermions and $\Nf\!=\!2$ or $\Nf\!=\!2\!+\!1$
violates flavour (or taste) symmetry and low energy unitarity, properties
our analysis relies on.
Under the assumption that these effects disappear in the continuum limit
\footnote{As of now, there is no proof, but rather a lively debate on this
issue in the literature \cite{squareroot}.}, the finite volume shift of an
observable like $\fpi/\fk$ can be calculated in staggered chiral perturbation
theory (this is what the MILC Collaboration does \cite{Aubin:2004fs}), but
one cannot enjoy the benefits of the L\"uscher formula.
We stress that no such conceptual issues arise with dynamical Wilson-type
fermions.
In such a case our continuum formulae can be directly applied to the data at
finite lattice spacing with cut-off effects bringing only mild (i.e.\
numerically irrelevant) modifications as discussed in App.\,\ref{app:cutoff}.


\subsection{Low energy constants from finite volume effects}

In the present section we discuss whether one can use these finite volume
effects to obtain information on the low energy constants from lattice
calculations.
At first sight this seems unpractical, because these effects are quite small
and decay exponentially with $M_\pi L$.
This means that, roughly speaking, if one wishes to obtain information on
(a combination of) low energy constants to a certain accuracy, one has to
calculate the corresponding particle mass or decay constant to an accuracy
which is about two orders of magnitude higher, and this is a challenge.
The asymptotic formulae provide a connection between finite volume effects on
two-point functions and (infinite volume) four-point functions, and thus give
us access to low energy constants that appear as local contributions to
four-point functions.
The question is then what alternative ways one has to measure these constants
on the lattice.
It is well known that the $P \pi$ scattering amplitudes which govern the finite
volume corrections to the mass of the $P$ particle can be obtained more
directly by evaluating the finite volume dependence of the energy of the
state of two particles $P$ and $\pi$ enclosed in a box \cite{Luscher:1986pf}.
This method is more direct because the effect is suppressed only by powers of
the volume rather than exponentially.
Still, such a calculation is very difficult, also because the typical volume
needed is quite large.

There is another important difference between the two methods. As we have
seen in Sect.~\ref{sec:simplified} the analytic representation of the
finite volume effects is enormously simplified if one Taylor expands the
amplitudes in the L\"uscher-type formulae. Here, two terms in the Taylor
expansion are enough for a very accurate representation. The extraction of
the low energy constants from these effects can be viewed as a two-step
process: one first determines the values of the first two Taylor
coefficients of the amplitudes at $\nu=0$, and then from these the low
energy constants. The chiral representation enters only in this second
step. Analogously, if one uses the method with two particles in a box, one
first determines the scattering lengths, and then extracts from these the
relevant low energy constants. As discussed in
\cite{Colangelo:2001df}, the scattering lengths have a badly converging
chiral expansion, such that only at very small quark masses one would be
able to reliably extract the low energy constants. This happens because one
is evaluating the amplitude on top of the threshold singularity. At
$\nu=0$, below threshold and away from any other singularity, the amplitude
displays a better convergence. 
For all these reasons we believe it is worthwhile to explore this
alternative route.

\begin{figure}
\centering
\includegraphics[width=14cm]{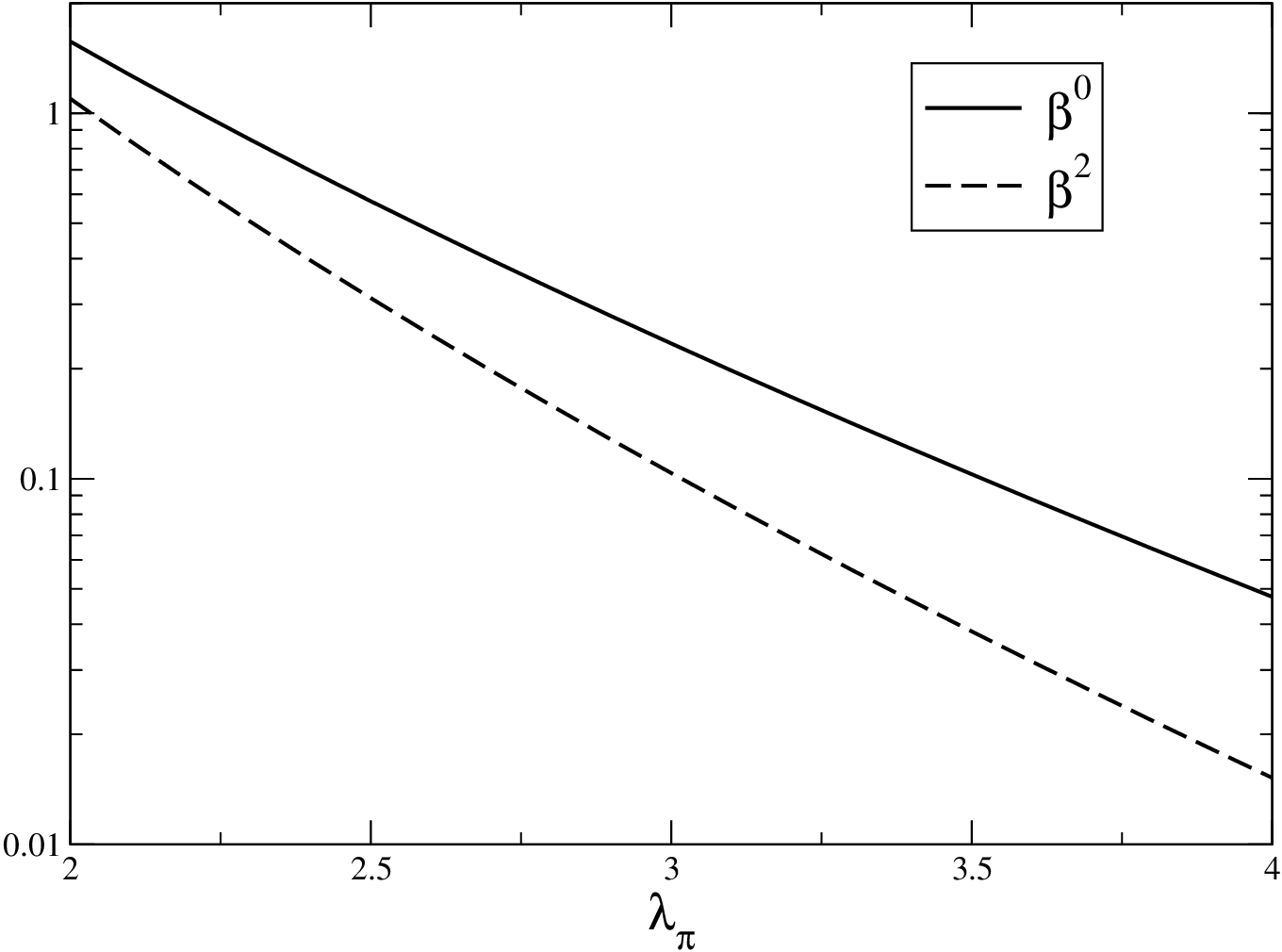}
\vspace{-2mm}
\caption{The functions $\beta^{0,2}$.}
\label{fig:betas}
\end{figure}

The quantities which are worth considering for our scope are $M_\pi$,
$F_\pi$ and $F_K$. We write the relative finite volume shifts in the form 
\be
R_X=R_X^0+c_X \Big( \beta_{}^0 L_X^0+\beta_{}^2 L_X^2 \Big)
\ee
where $R_X^0$ represents contributions independent of the low energy
constants, and the coefficients $c_X$ are defined as
\be
c_{M_\pi}= -\xi_\pi^2 \;,\quad 
c_{F_\pi}=  \xi_\pi^2 \;,\quad 
c_{F_K}=12 N \frac{F_\pi}{F_K} \xi_\pi \xi_K \fs
\ee
The functions $\beta^{0,2}$ are series of Bessel functions and depend on
$\lambda_\pi$ only 
\be
\beta^{0,2}=\sum_{n=1}^\infty
{m(n)\ovr\sqrt{n}\lambda_\pi}B^{0,2}(\sqrt{n}\la_\pi) \fs
\ee
They are plotted in Fig.~\ref{fig:betas}.
The $L_X^{0,2}$ are the combinations of low energy constants that appear in
each of the quantities to next-to-leading order. In particular we read off
from the formulae (\ref{RMP},\ref{RFP}) and
(\ref{explicit_Mpi},\ref{explicit_Fpi},\ref{explicit_Fk})
\bdm
\begin{array}{llllll}
L_{\mpi}^0&=&2(\bar\ell_1+{2\ovr3}\bar\ell_2)-{5\ovr4}\bar\ell_3-\bar\ell_4&
L_{\mpi}^2&=&-{4\ovr3} (\bar\ell_1+ 4 \bar\ell_2 )
\\[1.5mm]
L_{\fpi}^0&=&2(\bar\ell_1+{2\ovr3}\bar\ell_2)-3\bar\ell_4&
L_{\fpi}^2&=&-{8\ovr3} (\bar\ell_1+ 4 \bar\ell_2 )
\\[1.5mm]
L_{\fk }^0&=&  x_{\pi K}(4L_1^\mr{r}+L_3^\mr{r}-2L_4^\mr{r})
               -{1\ovr4}(1+x_{\pi K})L_5^\mr{r}\qquad&
L_{\fk }^2&=&-2x_{\pi K}(4L_2^\mr{r}+L_3^\mr{r})\;.
\end{array}
\edm
It is interesting that both $M_\pi$ and $F_\pi$ are sensitive to the same
combination of $\bar\ell_1$ and $\bar\ell_2$ (note that $\bar\ell_3$
and $\bar\ell_4$ can be pinned down directly from the quark mass 
dependence of $M_\pi$ and $F_\pi$, respectively).
This means that one can determine the low energy constants which fix the
$\pi\pi$ scattering amplitude also from the finite volume effects for
$F_\pi$ where the $\pi\pi$ scattering amplitude does not appear. One should
remark that the sensitivity to the combination
$(\bar\ell_1+{2\ovr3}\bar\ell_2)$ is the same, whereas $F_\pi$ is a factor
two more sensitive to the combination $(\bar\ell_1+4\bar\ell_2)$. 

To discuss the numerics we fix $M_\pi\!=\!300$ MeV and $L\!=\!2$ fm. In this
setting $\beta^0\!=\!0.22$, $\beta^2\!=\!0.10$, and
$c_{\mpi}=-0.0026$, $c_{\fpi}=0.0026$. A change of one unit in the combination
$(\bar\ell_1+{2\ovr3}\bar\ell_2)$ then generates a shift of about 0.12\% in
both $M_\pi(L)$ and $F_\pi(L)$. According to Tab.\,\ref{tab:SU2SU3}
this linear combination is known from phenomenology to be
\be
(\bar\ell_1+{2\ovr3} \bar\ell_2)_{|_{M_\pi=300 \mathrm{MeV}}}=-0.1\pm0.6 \fs 
\ee 
On the other hand a change of one unit in $(\bar\ell_1+ 4 \bar\ell_2)$ 
modifies $F_\pi(L)$ by 0.07\% and $M_\pi(L)$ by half that much. This linear
combination, however, is known to be larger from phenomenology
\be
(\bar\ell_1+ 4 \bar\ell_2)_{|_{M_\pi=300 \mathrm{MeV}}}= 9.1\pm0.5 \fs
\ee
This numerical example indicates that in order to get a reasonable account on
these combinations of low energy constants, one would have to control the pion
mass and decay constant for $M_\pi\!=\!300\MeV$ and $L\!=\!2\fm$ to less than 1
permille, which is a real challenge.

An alternative way, as already mentioned, would be to calculate the $\pi\pi$
$S$-wave $I\!=\!2$ scattering lengths via L\"uscher's method
\cite{Luscher:1986pf}.
There are some results with dynamical fermions for this quantity
\cite{Yamazaki:2004qb}, but not yet precise enough and for low enough pion
masses that would allow an extraction of the relevant low energy constant.
We mention in passing that both $a^2_0$ and $a^0_0$ (the latter is even more
difficult to calculate on the lattice, and shows a very badly converging
chiral expansion \cite{Colangelo:2001df}) are sensitive to the combination
$\bar\ell_1+2 \bar\ell_2$, and therefore provide complementary information
to the one that would be obtained from the finite volume effects we have
discussed here.

The numerics for the $F_K$ case is similar. Again, in order to get
a sensitivity comparable to the one which characterizes the
phenomenological determination (notice that the $L_i^\mr{r}$ constants
are usually given in units of $10^{-3}$) one would have to calculate $F_K$
to the permille accuracy.


\section{Conclusions}


In this paper we have carefully analyzed the finite volume effects on
masses and decay constants of the lightest pseudoscalar mesons. 
The theoretical framework in which we carry out this analysis has been set
up long ago by Gasser and Leutwyler \cite{Gasser:1987zq} and by L\"uscher
\cite{Luscher:1985dn}. The former two discussed how ChPT can be adapted to
the case of a finite box and then be used to calculate finite volume
effects. The latter author derived a formula valid for large volumes which
expresses the finite volume effects in terms of an integral over a physical
scattering amplitude. This formula does not rely on ChPT, but for QCD is
best used in combination with ChPT: this is the tool to provide a
representation of the necessary scattering amplitude as a function of quark
masses. More recently a formula {\em \`a la} L\"uscher for decay constants
has been derived by two of us \cite{Colangelo:2004xr}.  In the present
paper we have proposed a resummation of the asymptotic formulae as the best
tool to study finite volume effects for masses and decay constants. The
resummation is a simple, ``kinematical'' extension of the asymptotic
formulae, which however does improve the algebraic accuracy of the
formula. Used in combination with the chiral representation for the
scattering amplitude in the integral it yields an accurate determination of
finite volume effects.

Our numerical results show that decay constants and masses of the pseudoscalars
are in general little affected by the finite spatial size of the box
(as soon as the box is large enough that ChPT can be applied, $L\!\geq\!2\fm$).
For the smallest acceptable values of $M_\pi L$ (in the $p$-regime, in
which we are working, this quantity has to be larger than one) they are
typically of the order of a few percent for $M_\pi$, $F_\pi$ and $F_K$. We
have seen that in the $p$-regime $M_K$ and $M_\eta$ are practically
insensitive to the box size. Independently of the exact size of these
corrections we could always check the convergence of the chiral expansion
and conclude that these finite volume effects are under good theoretical
control. This means that if one's goal is to calculate masses or decay
constants one can use the results of this paper to choose the volume in
order to minimize the calculational costs. For example, by using a box of
$2\fm$ size (which our results show to be sufficiently large) and
explicitly correcting for the finite volume effects one can save
computational costs with respect to a $2.5\fm$ size box
which (for pion masses of about 300 MeV or larger) gives finite
volume effects below 1\%. The gain in CPU time that comes from such a reduction
of the volume by almost a factor 2 can be used more fruitfully by pushing
towards smaller lattice spacings and/or lighter quark masses.

Since our results are theory predictions, an explicit check by lattice
calculations would of course be very welcome, but it would require a
high precision. Once this level of accuracy will be reached, our
formulae will become particularly useful. First, in order to correct for
these effects in all applications which require a high precision, like the
evaluation of the $F_K/F_\pi$ ratio which, as suggested by Marciano
\cite{Marciano:2004uf}, leads to a determination of $V_{us}$
\cite{Aubin:2004fs}. Second, if one wants to use these finite size effects
as a means to determine low energy constants on the lattice. As we have
discussed, despite the fact that these effects are exponentially
suppressed and numerically quite small in all practical situations, they
do offer the advantage of involving in two-point functions low energy
constants which, otherwise, appear only in four-point functions. The latter are
quantities which are very difficult to determine directly on the lattice,
and it may therefore turn out to be easier to see them indirectly, through
a small correction to an ``easy'' quantity like the pion mass or decay
constant. 


\subsection*{Acknowledgments}

We are indebted to Peter Hasenfratz, Heiri Leutwyler and Rainer Sommer
for useful discussions and/or comments on the manuscript.
This work has been supported by the Schweizerischer Nationalfonds and
partly by the EU ``Euridice'' program under code HPRN-CT2002-00311.

\begin{appendix}


\section{The integrals $S_{M_P}^{(n)}$ and $S_{F_P}^{(n)}$}
\label{app:In}


In this appendix we give explicitly the contributions from loop functions
to the finite volume effects. These have been introduced and defined in
Sect.~\ref{sec:analytic}. The expression for the pion related integrals
have already been given in \cite{Colangelo:2003hf,Colangelo:2004xr}. We
give them here for convenience:
\bea
S^{(4)}_{M_\pi}&=&
{13\over3}R_0^0-{16\over3}R_0^1-{40\over3}R_0^2
\nonumber
\\
S_{\mpi}^{(6)}&=&
 R_0^0\Big({817\ovr27}+{80\ovr9}\lb_2-5\lb_3+{52\ovr3}\lb_4\Big)
-{2\ovr3}R_0^1\Big({313\ovr9}+{40\ovr3}\lb_2+32\lb_4\Big)
\nonumber\\
&+&R_0^2\Big({292\ovr27}-8\lb_1-{128\ovr9}\lb_2-{160\ovr3}\lb_4\Big)
+{4\ovr3}R_0^3\Big(-{47\ovr9}+4\lb_1-4\lb_2\Big)
\nonumber\\
&+&{1\ovr9}R_1^0\Big(1-{\pi^2\ovr2}\Big)
+{1\ovr9}R_1^1\Big(128-{\pi^2\ovr8}\Big)
-{1\ovr3}R_1^2\Big({100\ovr3}+{\pi^2\ovr8}\Big)
\nonumber\\
&+&{1\ovr6}R_2^0\Big(7-{\pi^2\ovr3}\Big)
+{1\ovr9}R_2^1\Big(16+{7\pi^2\ovr8}\Big)
+{\pi^2\ovr24}R_2^2
\nonumber\\
&-&{46\ovr9}R_3^0
-{32\ovr9}R_3^1
-{32\ovr3}R_3^2
+{40\ovr3}\Big(R_4^0+R_4^1\Big)
\nonumber
\\
S^{(4)}_{F_\pi}&=&
\frac{4}{3}\Big( R_0^0 - R_0^1 - 10 R_0^2 \Big) 
-{13\ovr6}R_0^{0\prime}+{8\ovr3}R_0^{1\prime}+{20\ovr3}R_0^{2\prime}
\eea
where the integrals $R_i^k$ are defined as 
\bea
R_0^{k(\prime)}\!&\!\equiv\!&\!R_0^{k(\prime)}(\sqrt{n}\la_\pi)=
\left\{{\mr{Re}\atop\mr{Im}}\right.
\int_{-\infty}^{\infty}\!d\til y\;\til y^k\,e^{-\sqrt{n(1+\til y^2)}\,\la_\pi}\,
g^{(\prime)}(2+2\ri\til y)\qquad\;\;\;\;
\mr{for}\;\left\{{k\;\mr{even}\atop k\;\mr{odd}}\right.
\label{eq:R0k}
\\
R_i^{k}\;\;\!&\!\equiv\!&\!\;\;R_i^{k\;}(\sqrt{n}\la_\pi)=
\left\{{\mr{Re}\atop\mr{Im}}\right.
\int_{-\infty}^{\infty}\!d\til y\;\til y^k\,e^{-\sqrt{n(1+\til y^2)}\,\la_\pi}\,
N^2\,K_i^{\pi\pi}(2+2\ri\til y)\;\;\;\;
\mr{for}\;\left\{{k\;\mr{even}\atop k\;\mr{odd}}\right.
\label{eq:Rik}
\eea
with
$g^{(\prime)}$ defined in (\ref{def_g_gprime}) and
\bea
K_1^{\pi\pi}(x)&=&{1\ovr  N^2\si^2}\Big[g(x)-2\Big]^2
\nonumber\\
K_2^{\pi\pi}(x)&=&{1\ovr  N^2}\Big[g(x)^2-4g(x)\Big]
\nonumber\\
K_3^{\pi\pi}(x)&=&{1\ovr 2N^2\si^4x}
\Big[2g(x)^3-12g(x)^2+24g(x)+2\pi^2\si^2g(x)-16-\pi^2\si^2x\Big]
\nonumber\\
K_4^{\pi\pi}(x)&=&{1\ovr \si^2x}
\Big[K_0^{\pi\pi}(x)+{1\ovr2}K_1^{\pi\pi}(x)+{1\ovr3}K_3^{\pi\pi}(x)
+{(\pi^2-6)x\ovr12N^2}\Big]
\eea
with $\si=\sqrt{1-4/x}$.
As seen in (\ref{eq:R0k}), in $S^{(4)}_{F_\pi}$ integrals over the
derivative of the function $g$ appear. This is a consequence of the
subtraction procedure (\ref{Eq:subFp}). In the practical implementation one
writes $s_1=2\Mpi^2-2\nu \Mpi+(Q^2-\Mpi^2)/2-s_3/2$ and
$s_2=2\Mpi^2+2\nu\Mpi+( Q^2-\Mpi^2)/2-s_3/2$ to trade $s_1,s_2$ for
$Q^2-\Mpi^2,\nu$ and expands $A_\pi^{I=0}$ consistently in these new
variables. For instance in $F_0$ one substitutes $\bar J(s_1)\to \bar
J(2\Mpi^2-2\nu\Mpi)+\bar J'(2\Mpi^2-2\nu\Mpi)[Q^2-\Mpi^2-s_3]/2$.

The loop integrals for kaon and eta finite volume effects read:
\begin{eqnarray}
  S_{\mk}^{(4)} &=& 3 \bigg\{
  \frac{3}{32} (1+x_{\pi K})^2  S^{0,1}_{K\pi}   
  -\frac{5}{8} (1+x_{\pi K}) S^{1,1}_{K\pi}
  -\frac{19}{8} S^{2,1}_{K\pi}
  -\frac{3}{16} (1-x_{\pi K}^2) S^{0,3}_{K\pi} \nn
  &+&\frac{13}{8} (1-x_{\pi K}) S^{1,3}_{K\pi}
  -\frac{3}{2} \left( x_{\pi K} S^{0,5}_{K\pi}+S^{2,5}_{K\pi} \right)
  + \frac{1}{96} (1+x_{\pi K})^2 S^{0,1}_{\eta K} \nn
  &-& \frac{1}{8} (1+x_{\pi K}) S^{1,1}_{\eta K}
  - \frac{3}{8} S^{2,1}_{\eta K} 
   + \frac{1}{16}(1+x_{\pi K})(3 x_{\eta K}+2 x_{\pi K}-5) S^{0,3}_{\eta K}
   \nn
  &+& \frac{3}{8} (1-2 x_{\pi K}+x_{\eta K}) S^{1,3}_{\eta K}
   -\frac{3}{2}\left( x_{\pi K} S^{0,5}_{\eta K}
  + S^{2,5}_{\eta K} \right)  \bigg\} \nn
  S_{F_K}^{(4)} &=& 
     \frac{\mk}{\mpi}\bigg[ -\frac{15}{16} (1+x_{\pi K}) S^{1,1}_{K\pi} 
     - \frac{57}{8} S^{2,1}_{K\pi} 
     - \frac{9}{64}(1 +x_{\pi K})^2 S^{0,2}_{K\pi}  \nn
     &+&\frac{15}{16}(1 +x_{\pi K}) S^{1,2}_{K\pi}
     + \frac{57}{16} S^{2,2}_{K\pi} 
     +\frac{9}{32} (1 -5 x_{\pi K}) S^{0,3}_{K\pi}
     + \left( 6 -\frac{15}{8}  x_{\pi K}\right ) S^{1,3}_{K\pi} \nn
     &+& \frac{15}{4} S^{2,3}_{K\pi}
     + \frac{9}{32} (1 -x_{\pi K}^2) S^{0,4}_{K\pi}
     - \frac{39}{16} (1 -x_{\pi K}) S^{1,4}_{K\pi} \nn
     &-& \frac{9}{4}\left(x_{\pi K} S^{0,5}_{K\pi}
     + 2 S^{2,5}_{K\pi} - x_{\pi K} S^{0,6}_{K\pi}
     - S^{2,6}_{K\pi} \right) \nn
   &-&\frac{3}{16} ( 1 +x_{\pi K} ) S^{1,1}_{\eta K}
     -\frac{9}{8} S^{2,1}_{\eta K}
     -\frac{1}{64} (1 +x_{\pi K})^2 S^{0,2}_{\eta K}
     + \frac{3}{16}(1 +x_{\pi K}) S^{1,2}_{\eta K} \nn
     &+& \frac{9}{16} S^{2,2}_{\eta K}
     +\left(\frac{27}{32} (x_{\eta K}-1)-\frac{3}{16}(1+x_{\pi K})\right)  
     S^{0,3}_{\eta K} \nn
   &+& \frac{3}{8}(4 - 2 x_{\pi K} + 3x_{\eta K} ) S^{1,3}_{\eta K}  
     +\frac{9}{4} S^{2,3}_{\eta K}
     + \frac{3}{32}(5 - 3 x_{\eta K}) (1-x_{\pi K}^2)
     S^{0,4}_{\eta K} \nn
   &-&\frac{9}{16}(1-2 x_{\pi K}+x_{\eta K}) S^{1,4}_{\eta K}
     -\frac{9}{4} \left( x_{\pi K} S^{0,5}_{\eta K}
      +2 S^{2,5}_{\eta K} - x_{\pi K} S^{0,6}_{\eta K}
      - S^{2,6}_{\eta K} \right) \bigg] \nn
S^{(4)}_{M_\eta}&=&  
              x_{\pi\eta}^2 T^{0,1}_{KK}
           -6 x_{\pi\eta} T^{1,1}_{KK}
           -9 T^{2,1}_{KK} 
+ \frac{2}{3}  x_{\pi\eta}^2 T^{0,1}_{\eta\pi} 
\end{eqnarray}
where we have introduced the following abbreviations
\beq
x_{PQ} = \frac{M_P^2}{M_Q^2}   \;,\qquad
\ell_P = \ln\left(\frac{M_P^2}{\mu^2}\right) \fs
\eeq
The integrals $B^{2k}$ are proportional to modified Bessel functions
\begin{equation}
B^{2k}\equiv B^{2k}(\sqrt{n}\la_\pi)
=\int_{-\infty}^{\infty}\!d\til y\;\til y^{2k}\,
e^{-\sqrt{n(1+\til y^2)}\,\la_\pi}
={\Gamma(k+1/2)\ovr\Gamma(3/2)}
\Big( {2\ovr\sqrt{n}\la_\pi} \Big)^k K_{k+1}(\sqrt{n}\la_\pi)
\end{equation}
and the quantities $S^{k,I}_{PQ}$ and $T^{k,I}_{PQ}$ are integrals
over functions $g_{PQ}^{I}$ which occur at 
one-loop order in the chiral expansion. They are all
analytical along the integration line. The expressions $S^{k,I}_{PQ}$
and $T^{k,I}_{PQ}$ are defined as
\begin{eqnarray}
 S_{PQ}^{k,I} &=& \left\{{\mr{Re}\atop\mr{Im}}\right. 
                  N x_{\pi K}^{(k+1)/2} \int_{-\infty}^{\infty} 
                  d\til y\;\til y^k\,
                  e^{-\sqrt{n(1+\til y^2)}\,\la_\pi}
                  g_{PQ}^{(I)}(\mk^2+\mpi^2+2\ri \mk\mpi\til y)
                  \qquad\mr{for} 
                  \left\{ {k \; \mr{even}}\atop 
                          {k \; \mr{odd}} \right.\nn
T_{PQ}^{k,I} &=& \left\{{\mr{Re}\atop\mr{Im}}\right. 
                  N x_{\pi\eta}^{(k+1)/2} \int_{-\infty}^{\infty} 
                  d\til y\;\til y^k\,
                  e^{-\sqrt{n(1+\til y^2)}\,\la_\pi} 
                  g_{PQ}^{(I)}(\,\me^2\,+\mpi^2+2\ri \,\me\mpi\til y)
                  \qquad\mr{for} 
                  \left\{ {k \; \mr{even}}\atop 
                          {k \; \mr{odd}} \right.\nonumber
\end{eqnarray}
with
\bdm
\begin{array}{rclrcl}
g_{PQ}^{(1)}(x) \!&\!=\!&\! \bar{J}_{PQ}(x)\;,\qquad&
g_{PQ}^{(2)}(x) \!&\!=\!&\! \mk^2 \bar{J}'_{PQ}(x)\qquad
\\
g_{PQ}^{(3)}(x) \!&\!=\!&\! K_{PQ}(x)\;,\qquad&
g_{PQ}^{(4)}(x) \!&\!=\!&\! \mk^2 K'_{PQ}(x)\qquad
\\
g_{PQ}^{(5)}(x) \!&\!=\!&\! \bar{M}_{PQ}(x)\;,\qquad&
g_{PQ}^{(6)}(x) \!&\!=\!&\! \mk^2 \bar{M}'_{PQ}(x)\;.
\end{array}
\edm
For completeness, we give the explicit expressions for the
$g_{PQ}^{I}$ \cite{Bijnens:1994ie}. All functions can be expressed
in terms of the subtracted scalar integral $\bar{J}(t)\!=\!J(t)\!-\!J(0)$
evaluated in four dimensions
\begin{equation}
J(t) = -\ri
\int\!\frac{d^dp}{(2\pi)^d}\;\;\frac{1}{((p+k)^2 - M^2)(p^2 - m^2)}
\end{equation}
with $t = k^2$.
The functions used in the text are then
\begin{eqnarray}
\label{Eq.:loop_functions}
\bar{J}(t)&=&-\frac{1}{N}\int_0^1 dx~
\ln\frac{M^2 - t x(1-x) - \Delta x}{M^2 - \Delta x}
\nn&=&
\frac{1}{2 N}\left\{
2 + \frac{\Delta}{t}\ln\frac{m^2}{M^2} -\frac{\Sigma}{\Delta}
\ln\frac{m^2}{M^2} - \frac{\sqrt{\rho}}{t}
\ln\frac{(t+\sqrt{\rho})^2 -
\Delta^2}{(t-\sqrt{\rho})^2-\Delta^2}\right\}
\nn
\bar{J}'(t) &=& -\frac{2}{N} \frac{M^2
  m^2}{(t-\Sigma)^2-\rho}\frac{1}{t^2} \bigg[ 2t + \Delta\ln\frac{m^2}{M^2}
   +\frac{t \Sigma-\Delta^2}{\sqrt{\rho}} \ln\frac{(t+\sqrt{\rho})^2 -
\Delta^2}{(t-\sqrt{\rho})^2-\Delta^2} \bigg]\nn 
K(t)&=&\frac{\Delta}{2t}\bar{J}(t)
\nn
K'(t) &=& -\frac{\Delta}{2t}\left( \frac{\bar{J}(t)}{t}-\bar{J}'(t) \right)\nn 
\bar M(t) &=& \frac{1}{12t}\left\{ t - 2 \Sigma \right\} \bar{J}(t)
+ \frac{\Delta^2}{3 t^2} \bar{J}(t)
+ \frac{1}{18 N} - \frac{1}{6 N t} \left\{
      \Sigma + 2 \frac{M^2 m^2}{\Delta}
     \ln\frac{m^2}{M^2} \right\}
\nn
\bar M'(t) &=& \frac{1}{6 t^2}\bigg[ \frac{\Sigma t-4\Delta^2}{t} \bar{J}(t)
               +\frac{1}{2}(t^2 - 2t\Sigma + 4\Delta^2) \bar{J}'(t)
+ \frac{1}{N} \left(\Sigma + \frac{2 M^2 m^2}{\Delta} \ln
  \frac{m^2}{M^2}  \right)
                              \bigg]
\end{eqnarray}
where
\beq
\Delta = M^2 - m^2\;,\qquad
\Sigma = M^2 + m^2\;,\qquad
\rho   = \rho(t,M^2,m^2) = (t+\Delta)^2 - 4tM^2 \fs
\eeq
In the text these are used with subscripts
\begin{equation}
\bar{J}_{PQ}(t) = \bar{J}(t)~~~\mbox{with}~~~M = M_P , m = M_Q~
\end{equation}
and similarly for the other symbols. 
We add a remark concerning the analyticity properties of the loop
functions. The asymptotic formula requires them to be evaluated for 
complex arguments. There is one case, where the
representation of Eq.(\ref{Eq.:loop_functions}) does not yet provide an
unambiguous analytic continuation, namely for 
$\bar J_{P Q}(M_P^2+M_Q^2+2\ri M_P M_Q\til y)$, because
$\rho=-4M_P^2M_Q^2(1+\til y^2)$. The correct analytical continuation is given by
\begin{equation}
\sqrt{\rho} = 2\ri M_P M_Q \omega \nn
\end{equation}
with $\omega = \sqrt{1+\til y^2}$, 
implying for the logarithm in Eq.(\ref{Eq.:loop_functions}) 
(for $t=M_P^2+M_Q^2+2 \ri M_P M_Q \til y$), 
\begin{equation}
\ln\frac{(t+\sqrt{\rho})^2 -
\Delta^2}{(t-\sqrt{\rho})^2-\Delta^2} = 
\ln{\omega+\til y\ovr\omega-\til y}
\left\{
{+\ri\pi\qquad\mr{for}\;{\til y<0}\atop-\ri\pi\qquad\mr{for}\,{\til y>0}}
\right.
\fs 
\end{equation}
All loop functions are now well defined along the integration line in the
asymptotic formulae. 


\section{Cut-off effects}
\label{app:cutoff}


In this appendix we wish to discuss whether it is sufficient to calculate
finite volume effects in \emph{continuum} ChPT or whether cut-off effects
should be taken care of when correcting
\footnote{Here we assume that the finite volume correction is applied
\emph{before} the continuum and chiral extrapolations, thus interchanging
steps ($i$) and ($ii$) of Sect.\,1.
In practice such a change is helpful, since otherwise the volumes would have to
be matched to perform a well-defined continuum extrapolation and this would
require a priori knowledge of the lattice spacing that will come out of the
simulation.}
actual lattice data for the effect of
the finite spatial box length $L$.

Naive reasoning suggests that -- because finite volume effects are due to the
pion cloud around a particle and thus to pure IR physics -- such shifts
will be rather insensitive to the UV properties of the theory.
This is what one expects to hold as long as the cut-off is large compared to
the scale of chiral symmetry breaking, $\Lambda_\mr{XSB}\!\simeq\!1\GeV$.
With a lattice regularization all momenta are cut off at $\pi/a$, and with a
standard lattice spacing $a\!\simeq\!0.1\fm$ the resulting scale $\sim\!6\GeV$
is indeed much bigger than $\Lambda_\mr{XSB}$.

This intuitive argument can be refined in two ways.
The first option is to invoke an extension of ChPT designed to take care of
the effects of the finite lattice spacing $a$.
Of course, the details of this theory need to be tailored to the action used,
but generically the new Lagrangian follows from the old one by replacing the
low energy constants, e.g.\ $\ell_3\to \ell_3+\mr{const}\,w_3$ -- see
Ref.\,\cite{Bar:2004xp} for a recent review.
In consequence, the $O(p^4)$ formula (\ref{mpi}) takes the form
\beq
\Mpi(a,L)=M\,
\Big\{
1-{1\ovr4}x
\Big[\til\ell_3+\mr{const}\,\til w_3-\log(x)\Big]+{1\ovr2\Nf}x\til g_1(\la_\pi)
+O(x^2)
\Big\}
\eeq
where $x\!=\!M^2/(4\pi F)^2$, $M\!=\!\sqrt{2Bm}$ and
$\til\ell_3=\log(\Lambda_3^2/(4\pi F)^2)$.
In other words the very effect of such an extension concerns the particle
mass in \emph{infinite} volume, the fractional finite-size effect gets modified
by $a$-effects only at $O(p^4)$ in the chiral counting, viz.
\beq
\Mpi(a,L)=\Mpi(a)\,\Big(1+{1\ovr2\Nf}x\til g_1(\la_\pi) +O(x^2)\Big)
\;.
\eeq

The second option is to compare the generic one loop finite volume shift in the
continuum to a version in which the pion propagator is discretized in the
simplest 
\footnote{By considering (\ref{shift_latt_def},\ref{shift_latt_end}) we do
not indicate that this would yield a better estimate of the finite size
effects than the continuum form (\ref{shift_cont_def},\ref{shift_cont_end}).
The actual discretization of quarks and gluons will not lead to a simple
pion propagator, but we are interested in the absolute difference
$|g_1(M,L,0)\!-\!g_1(M,L,a)|$, since it is expected to correctly indicate the
order of magnitude of discretization effects in the finite volume shift of
actual data. We stress that our discretization is similar in spirit, but
not identical to the one used in Ref.\,\cite{Borasoy:2004zf}.}
possible way
\bea
g_1(M,L,0)&=&
\int\!{dp_0\ovr2\pi}\;\bigg\{
{1\ovr L^3}\sum_{{2\pi\ovr L}\mb{Z}^3}\;
{1\ovr \mb{p}^2+p_0^2+M^2}
-\,
\int\!{d^3\mb{p}\ovr(2\pi)^3}\;
{1\ovr \mb{p}^2+p_0^2+M^2}
\bigg\}
\label{shift_cont_def}
\\
g_1(M,L,a)&=&
\int\!{dp_0\ovr2\pi}\;\bigg\{
{1\ovr L^3}\sum_\mr{ finite }\;
{1\ovr \hat{\mb{p}}^2+p_0^2+M^2}
-\!
\int\limits_{-\pi/a }^{+\pi/a }\!\!{d^3\mb{p}\ovr(2\pi)^3}\;
{1\ovr \hat{\mb{p}}^2+p_0^2+M^2}
\bigg\}
\label{shift_latt_def}
\eea
[the definition $\hat{\mb{p}}^2\!=\!{4\ovr a^2}\sum_i\sin^2({a\ovr2}p_i)$ has
been used and the finite sum runs over $p_i={2\pi\ovr L}n_i$ with
$n_i\!\in\!\{0,...,N\!-\!1\}$ and $N\!=\!L/a$ an integer] and verify that
the difference is small compared to the shift itself, i.e.\
$|g_1(M,L,0)\!-\!g_1(M,L,a)|\!\ll\!g_1(M,L,0)$ for standard values of $M,L,a$.
With
\bea
g_1(M,L,0)\!&\!=\!&\!
\int\limits_0^\infty\!dt\;
{1\ovr16\pi^2 t^2}\sum\limits_{n\geq1}
m(n)\,e^{-nL^2/(4t)-M^2t}\,\,
=
{1\ovr4\pi^2}\sum\limits_{n\geq1}
m(n){M\,K_1(\sqrt{n}ML)\ovr\sqrt{n}L}
\label{shift_cont_end}
\\
g_1(M,L,a)\!&\!=\!&\!
\int\limits_0^\infty\!dt\;
\Bigg(
\bigg[{1\ovr L}\sum\limits_{n=0}^{N-1}
\exp\Big(-{4t\ovr a^2}\sin^2({\pi n\ovr N})\Big)\bigg]^3
-
\bigg[{I_0({2t/a^2})\ovr a\,e^{{2t/a^2}}}\bigg]^3
\Bigg)\,
{e^{-tM^2}\ovr\sqrt{4\pi t}}
\label{shift_latt_end}
\eea
[cf.\ Tab.\,\ref{tab:multiplicities} for $m(n)$]
and $M\!=\!300\MeV, L\!=\!2\fm, a\!=\!0.1\fm$ one finds
$(4\pi/M)^2g_1(M,L,0)\!=\!0.4374$ and $(4\pi/M)^2g_1(M,L,a)\!=\!0.4400$, thus
a difference of less than a percent in a quantity designed to correct actual
data by --~at most~-- a few percent.
It is hence sufficient to calculate the finite volume effects in continuum
ChPT.
This conclusion was also reached in Ref.\,\cite{Sharpe:1992ft}.


\section{Effects due to kaon and eta loops}
\label{app:KE}


Our formulae (\ref{RMP}, \ref{RFP}) take only the effects due to virtual pion
loops into account.
In other words, they neglect the contribution to $M_P(L)\!-\!M_P$ and
$F_P(L)\!-\!F_P$ coming from kaon and eta loops ``around the world''.
In this appendix we want to discuss to which extent this is justified.

Consider a soon-to-be standard $\Nf\!=\!2\!+\!1$ simulation with
$\Mpi\!=\!300\MeV$ and the strange quark fixed at its physical value and
$\mk\!=\!530\MeV$ in consequence (see Fig.\,\ref{fig:fpifkmkme}).
With a box size $L\!=\!2\fm$ a first crude estimate says that the kaon loop
effects will be down, relative to the pion loops, by a factor
$e^{-(\mk-\mpi)L}=0.1$, and a 10\% correction on the fractional
finite volume effect is not exactly small.
However, this correction should be compared to the absolute size of
the effect and the statistical error
and from Tabs.\,\ref{tab:Rmpi}-\ref{tab:Rfpi} and/or
Figs.\,\ref{fig:Rmpi}-\ref{fig:Rmkaet} it follows that it is safe to neglect
such a correction at the permille level.
Finally, we mention that one expects NNLO contributions for $R_{\fpi},R_{\fk}$
to be of the same order of magnitude as for $R_{\mpi}$ and this means that
an additional pion loop could prove more important than a single kaon loop in
finite volume.

We have verified that kaon and eta loops ``around the world'' prove numerically
insignificant by comparing the pion-loop contributions to those of the kaons
and etas in the full 1-loop expressions calculated in $SU(3)$ ChPT in finite
volume. We find
\bea
R_{\mpi}&=&
{1\ovr4}\xi_\pi  \til g_1(\la_\pi)-{1\ovr12}\xi_\eta \til g_1(\la_\eta)
\label{Eq.:Rm}
\\
R_{\mk} &=&
{1\ovr6}\xi_\eta \til g_1(\la_\eta)
\label{RMka}
\\
R_{\me}&=&
{1\ovr2}\xi_K    \til g_1(\la_K)-{1\ovr3}\xi_\eta \til g_1(\la_\eta)
+{\Mpi^2\ovr M_\et^2}\Big[
-{1\ovr 4}\xi_\pi \til g_1(\la_\pi)
+{1\ovr 6}\xi_K   \til g_1(\la_K)
+{1\ovr12}\xi_\eta\til g_1(\la_\eta)\Big]
\label{RMet}
\eea
where $\la_P$ and $\til g_1$ have been defined in (\ref{def_la_p}) and
(\ref{g1til}), respectively, and
\bea
R_{\fpi}&=&        -\xi_\pi \til g_1(\la_\pi)
           -{1\ovr2}\xi_K   \til g_1(\la_K)
\\
R_{\fk} &=&-{3\ovr8}\xi_\pi \til g_1(\la_\pi)
           -{3\ovr4}\xi_K   \til g_1(\la_K)
           -{3\ovr8}\xi_\eta\til g_1(\la_\eta)
\label{check_bercirevic}
\\
R_{\fe} &=&-{3\ovr2}\xi_K   \til g_1(\la_K)
\;.
\label{Eq.:Rf}
\eea
These formulae deserve a few comments.
First, a few elementary checks: Our (\ref{Eq.:Rm}), (\ref{RMka}) and
(\ref{check_bercirevic}) 
agree with $\Delta\Mpi/\Mpi,\Delta\mk/\mk,\Delta\fk/\fk$ as given in
\cite{Becirevic:2003wk,Descotes-Genon:2004iu} and both the $R_M$ and the $R_F$
become degenerate in the $SU(3)$ limit ($m_u=m_d=m_s$), and furthermore, these
degenerate expressions agree with the result by Gasser and Leutwyler
\cite{Gasser:1986vb}, eqns.\,(\ref{mpi}) and (\ref{fpi}), specified to
$\Nf\!=\!3$.
Second, $R_{\mk}$ and $R_{\fe}$ have no $\til g(\la_\pi)$ (in other words
only kaon- and eta-loops contribute at one-loop order) and in $R_{\me}$ the
one-pion-loop contribution is suppressed by an extra factor $\Mpi^2/\me^2$.
Therefore, one expects $\mk(L)-\mk$, $\me(L)-\me$ and $\fe(L)-\fe$ to be small,
and the numerical investigation in Sect.\,6 specifies to which extent
this is true.
The numerical discussion also shows that, for a substantial range of pion
masses, we have $\me(L)<\me$, in (apparent) contradiction to our statement
in Sect.\,2 that finite volume effects will \emph{lift} the masses of the
pseudo Goldstone bosons. However, this just indicates that the general rule
may be overwhelmed by $SU(3)$ breaking effects.
As our formulae show, the $SU(3)$ breaking effects are accidentally
dominating (at one-loop level) only in the $R_{M_P}$ and not in the
$R_{F_P}$, i.e.\ $R_{F_P}\!<\!0$ for $P=\pi,K,\et$.
Finally, we wish to elaborate on a point already raised in the work by
Becirevic and Villadoro \cite{Becirevic:2003wk}.
The main message of the one-loop formulae (\ref{Eq.:Rm} - \ref{Eq.:Rf}) is
that finite volume effects and chiral logs are intimately related.
In this particular case, where there is only a tadpole contribution, the finite
volume shift follows from the quark mass dependence by the simple substitution
$\log(M_P^2/\mu^2)\to\log(M_P^2/\mu^2)+\til g_1(\la_P)$.
In practice, this means that one cannot extract ``chiral logs'' and the
pertinent low energy constants without controlling the finite volume effects.

\end{appendix}


\end{document}